\documentclass[11pt,a4paper]{article}
 
\usepackage{amsmath,amsthm,amssymb}
\usepackage{graphics,graphicx}
\usepackage{epsfig}
\usepackage{multicol}
\usepackage{color}
\makeatletter
\@addtoreset{equation}{section}
\makeatother

\begin{document}

\begin{flushright}
{RESCEU-37/12}
\end{flushright}
\begin{center}
\LARGE{\bf Cosmic strings with twisted magnetic flux lines and wound-strings in extra dimensions}
\end{center}

\begin{center}
\large{\bf Matthew Lake} ${}^{a}$\footnote{lake@resceu.s.u-tokyo.ac.jp}\large{\bf and Jun'ichi Yokoyama} ${}^{a,b}$\footnote{yokoyama@resceu.s.u-tokyo.ac.jp} 
\end{center}
\begin{center}
\emph{ ${}^a$ Research Center for the Early Universe \\ School of Science, The University of Tokyo \\ 7-3-1 Hongo, Bunkyo-ku, Tokyo 113-0033, Japan \\}
\emph{ ${}^b$ Kavli IPMU (WPI), The University of Tokyo \\ Kashiwa, Chiba 277-8582, Japan \\}
\vspace{0.1cm}
\end{center}


\begin{abstract}
We consider a generalization of the Nielsen-Olesen ansatz, in an abelian-Higgs model with externally coupled charge, which describes strings with twisted magnetic flux lines in the vortex core. The solution does not possess cylindrical symmetry, which leads to the existence of components of conserved momentum, both around the core-axis and along the length of the string. In addition, we consider a model of $F$-strings with rotating, geodesic windings in the compact space of the Klebanov-Strassler geometry and determine matching conditions which ensure energy and momentum conservation when loops chop off from the long-string network. We find that the expressions for the constants of motion, which determine the macroscopic string dynamics, can be made to coincide with those for the twisted flux line string, suggesting that extra-dimensional effects for $F$-strings may be mimicked by field-theoretic structure in topological defects. 
\end{abstract}


\section{Introduction} \label{Introduction}
Previous studies of cosmic strings in models with extra dimensions have, not unreasonably, focussed on the differences between the behavior of loops and networks in higher-dimensional space, compared to their dynamics in ``ordinary" four-dimensional spacetime. In particular, the intercommuting probability, $P$, may be greatly reduced in the former scenario, leading to the failure of the network to reach a scaling regime \cite{Copeland:2005,Avgoustidis:2004} and resulting in reduced loop production. Loop dynamics can also be significantly altered by the motion of the string in the compact space. While, for backgrounds with non-simply connected internal manifolds, windings are topologically stable \cite{Avgoustidis:2005}, in models with simply connected internal spaces this is not so, and windings must be stabilized dynamically by the existence of angular momentum in the compact directions. 
\\ \indent
In \cite{BlancoPillado} the criterion for the stability of geodesic windings for strings lying at the tip of the warped throat in the Klebanov-Strassler (KS) geometry \cite{KS} (corresponding to strings formed at the end of brane inflation \cite{Tye1}-\cite{Pogosian_Obs1}), was determined, and its implications for the dynamics of string loops were investigated in \cite{Cycloops}. Three distinct dynamical regimes were found to exist, depending on the initial value of the momentum in the angular directions of the $S^3$ which regularizes the conifold. Equivalently, these solutions could be classified according to the initial fraction of the total string length contained in the extra dimensions, labelled $\omega_{l}^2(t_i)$ in \cite{Cycloops}, where $t_i$ is the time of loop production. For $\omega_{l}^2(t_i)<1/2$, the loop was found to collapse to some critical (but nonzero) radius before re-expanding to its initial size and continuing to oscillate between these two values. For $\omega_{l}^2(t_i)>1/2$ an initial period of expansion is followed by oscillation between the initial configuration and some maximum radius. In the critical case, $\omega_{l}^2(t_i)=1/2$, a perfectly circular loop was found to be dynamically stabilized against oscillation by the exact balance of the string tension by the angular momentum. In the oscillatory cases, the period of oscillation was also characterized in terms of $\omega_{l}^2(t_i)$. 
\\ \indent
Such behavior contrasts strongly with the expected dynamics of strings in $(3+1)$ dimensions. The key difference is that wound-strings posses ``internal structure" from an effective four-dimensional perspective (though the Lorentz invariance of all physical quantities is preserved for boosts along the direction of the string in the full $(3+d+1)$-dimensional space).  For example, suppose that the string adopts a helical configuration and rotates around the compact space. In this case, the embedding resembles the motion of a screw-thread and results in translational motion of the string configuration parallel to itself in the infinite directions. Since this necessarily involves introducing a component of motion perpendicular to the direction of the string in the internal space, a momentum is generated with nonzero components in both the compact and infinite directions. By contrast, the translation of an unwound string ``parallel to itself" is physically meaningless. Regardless of the theoretical origin of the string (i.e. regardless of whether the we consider cosmic superstrings or topological defects), few string species investigated so far contain genuine internal structure along the direction of their length. By definition, in the absence of world-sheet (WS) fluxes, cosmic $F$- and $D$-strings \cite{Witten:1985,Polchinski_Intro} contain no internal structure whatsoever, and the topologically nontrivial structure of field-theoretic strings typically obeys cylindrical symmetry about the central string axis. 
\\ \indent
In this paper, we propose an ansatz for a string which contains helical lines of magnetic flux that destroy cylindrical symmetry. Associated with these twisted flux-lines are helical lines of constant phase which rotate around the central axis, in much the same way that wound-strings rotate around the compact space in extra-dimensional models. This leads to the existence of two winding numbers; the topological winding number, $|n| \geq 1$, which signals the existence of confined flux, and hence the string itself, and a nontopological winding number, $|m| \geq 0$, which characterizes the number of twists in the magnetic field lines (or equivalently, in the phase as we move along a line of constant $r$ and $\theta$ within the string core). We also compare the values of the energy and momenta obtained for this model, and the resulting loop dynamics, with those obtained for the wound $F$-strings considered in \cite{BlancoPillado,Cycloops}. For a simple choice of model parameters we find good agreement between the two, which suggests, as mentioned in the abstract, that extra-dimensional effects for cosmic superstrings may be mimicked by field-theoretic structure in topological defects.
\\ \indent
However, a key feature of the field-theoretic solution is that it requires the existence of additional charged matter, which couples to electromagnetic sector. A small amount of externally-coupled charge is required to regularize the solution for the gauge potential close to the string central axis, ensuring $A_{\mu}(r) \rightarrow 0$ as $r \rightarrow 0$. Without this additional matter, the gauge field diverges, leading to divergent energy, $E$, and total charge, $Q$. Interestingly though, the motion of this ``external" charged matter may make negligible contributions to the constants of motion, with the overwhelming majority of the nontrivial internal structure coming from the ``vacuum part" of the solution; that is, from the configuration of the usual abelian-Higgs fields. The additional matter is required only in the sense that, via its interaction with the gauge field, it allows the latter to be regularized close to $r=0$, even in the presence of twists in the field lines.
\\ \indent
For this reason, the solution presented here is not a true vacuum solution but rather an example of a hybrid-type string that may form initially if the abelian-Higgs fields undergo a phase transition in the presence of external charged matter. For other, more complex, Lagrangians the possibility remains that the gauge field contribution may be regularized purely by the presence of internally generated charges. If so, true vacuum strings which display apparent ``higher dimensional" dynamics may be formed via a similar mechanism to that considered in the following, hybrid, case.
\\ \indent
The structure of this paper is then as follows: In Sect. 2, we outline our model of wound-strings in the KS geometry, beginning with long, straight $F$-strings, before moving onto loops. Sect. 3 gives a general outline of the abelian-Higgs model (in the presence of additional charged matter), and the field configuration for long, straight, topological defect strings with twisted flux-lines is presented in Sect. 4, together with the calculation of the string energy and momenta. Sect. 5 describes the transition to loops in the field theory, using conserved quantities to determine the macroscopic loop dynamics. Based on the results of previous sections, a correspondence between the field theory and string theory parameters is suggested in Sect. 6 and specific models for realizing the field/string configurations considered here are briefly reviewed. Finally, the stability analysis for the wound-string is presented in Sect. 7, together with analogue results for the twisted-flux string case, based on the correspondences proposed in Sect. 6. Crucially, it remains necessary to perform a separate analysis for the field-theoretic string, in order to verify/determine the stability of the configuration, though this lies beyond the scope of the current paper. A brief review and discussion of our results and proposals for future work are given in Sect. 7.  
\\ \indent
The appendix sets the twisted flux line solution in the context of the existing literature, in particular, its relation to the traveling wave solutions for Nielsen-Olesen type strings \cite{Vachaspati&Vachaspati,Garfinkle&Vachaspati}. As discussed therein, the key physical difference between the two is that the traveling wave formalism  describes waves {\it on} strings, be they of zero-thickness or finite width, whereas the field-theoretic model presented here may be said to describe waves {\it within} strings. That is, in the former, energy and momentum are carried by the string due to the temporally and spatially-dependent displacement of its central axis whereas, in the latter, the string axis remains fixed (according to its general embedding, i.e. here a straight line or a loop), while the field configurations within the string core itself undergo temporal and spatial evolution. The evolution of this  ``internal" structure is therefore rightly the analogue of $F$-string evolution in the ``internal" space in string theory models, since it is the only way of introducing additional degrees of freedom beyond those associated with the movement  of the string in four-dimensional spacetime. 
\\ \indent
In principle, these two, physically different, types of field evolution may coexist for defect strings, so that the evolution of the internal structure of the string and the embedding of the central axis in the background space undergo mutual interaction. It is an interesting open problem to consider the effects of this interaction, though no such analysis is attempted here for the sake of brevity and clarity, due to the the likely complexity of the resulting equations of motion (EOM) for the abelian-Higgs fields.
\section{Wound-strings in the KS geometry} \label{Wound-strings in the KS geometry}
Using the Hopf vibration of the three-sphere \cite{HopfMap}, the metric in the KS geometry \cite{KS} may be written as
\begin{eqnarray} \label{KSmetric}
ds^2 = a^2g_{\mu\nu}dx^{\mu}dx^{\nu} - R^2[d\psi^2 + \sin^2(\psi)d\chi^2 + \cos^2(\psi)d\phi^2],
\end{eqnarray}
close to the warped-throat tip. Here $g_{\mu\nu}=\eta_{\mu\nu}$ is the Minkowski metric with signature $(+---)$, $0 < a <1$ is the warp factor induced by the back-reaction of the fluxes which stabilize the compact space, $R$ is the radius of the $S^3$, $\psi \in [0,\pi)$ is the polar angle and $\chi,\phi \in [0,2\pi)$ are the two azimuthal angles. The appropriate action for an $F$-string carrying no additional world-sheet (WS), fluxes is the Nambu-Goto action \cite{Nambu,Goto}
\begin{eqnarray} \label{NambuGoto}
S = -\mathcal{T}\int d^2\zeta \sqrt{-\gamma}, 
\end{eqnarray}
where $\gamma$ is the determinant of the induced metric on the WS
\begin{eqnarray} \label{indmet}
\gamma_{ab}(\tau,\sigma) = g_{MN}(\tau,\sigma)\frac{\partial X^{M}(\tau,\sigma)}{\partial \zeta^{a}}\frac{\partial X^{N}(\tau,\sigma)}{\partial \zeta^{b}},
\end{eqnarray}
$M,N \in \left\{0,1,2,...9\right\}$, $a,b \in \left\{0,1\right\}$, $\zeta^{0}=\tau,\zeta^{1}=\sigma$, which is equal to the WS area, and
\begin{eqnarray} \label{T}
\mathcal{T} = \frac{1}{2\pi \alpha'}
\end{eqnarray}
is the fundamental tension of the $F$-string, where $\alpha'$ is the Regge slope parameter and the fundamental string length-scale is given by
\begin{eqnarray} \label{ls}
l_{st} = \sqrt{\alpha'},
\end{eqnarray}
with $\hbar = c = 1$. The string energy-momentum tensor is 
\begin{eqnarray} \label{EMtens}
{}^{(s)}T^{M}{}_{N} = \frac{\partial \mathcal{L}}{\partial (\partial_M X^I)}\partial_{N}X^I - \delta^{M}{}_{N}\mathcal{L} = \frac{-2}{\sqrt{-g}}g_{IN}\frac{\delta S}{\delta g_{MI}},
\end{eqnarray}
where $\mathcal{L}=-\mathcal{T}\sqrt{-\gamma}$, and the conserved charges, ${}^{(s)}\Pi_J$, are
\begin{eqnarray} \label{Cons}
{}^{(s)}\Pi_{J} = \int d^3x \sqrt{-g} {}^{(s)}T^{0}{}_{J}.
\end{eqnarray}
These have dimensions $[l]^{-1}$ when $J=0$, or when $J=i$, where $x^i$ corresponds to a rectilinear coordinate, and are dimensionless when $J=j$, where $x^j$ represents an angular coordinate. The corresponding contravariant components, defined via
\begin{eqnarray} \label{Contra}
{}^{(s)}\Pi^{J} = \int d^3x \sqrt{-g} {}^{(s)}T^{0J},
\end{eqnarray}
therefore have units of energy or linear momentum, $[E]=[l]^{-1}$, for $J\in \left\{0,i\right\}$, and units of angular momentum, $[l]^{-2}$, for $J=j$. However, since ${}^{(s)}T^{0J}(\tau,\sigma)=g^{JI}(\tau,\sigma){}^{(s)}T^{0}{}_{I}(\tau,\sigma)$, that is, since the metric is generally a function of the string embedding, which may be time-dependent, the latter are not necessarily conserved directly. In this case, we may find the value of the conserved angular momentum by evalualting ${}^{(s)}T^{0J}(\tau_i,\sigma)$, where $\tau_i$ corresponds to the value of the WS time-coordinate at which we impose our initial conditions for the system. 
\\ \indent
In this paper, we use the notation ${}^{(s)}\Pi_{J}$ only when the nature of the conserved charge remains unspecified. To avoid confusion, we will use ${}^{(s)}P$, with an appropriate index,  to refer to quantities related to time-translation and rectilinear space-translation invariance, and ${}^{(s)}l$, with an angular-coordinate index, to refer to quantities related to rotational invariance. In addition, since we are working in warped space, it is important to include the correct factors from the metric when combining covariant and contravariant vectors. With this in mind, we define the total energy, $E_s$, linear momentum, $P_s$ and ``angular momentum", $l_s$, of the system via
\begin{eqnarray} \label{E,P,l}
E_s = \sqrt{{}^{(s)}P_0{}^{(s)}P^{0}}, \ \ \ P_s=\sqrt{-{}^{(s)}P_{i}{}^{(s)}P^{i}}, \ \ \ l_s=\sqrt{-{}^{(s)}l_{j}{}^{(s)}l^{j}(t_i)},
\end{eqnarray}
though, in the latter case, strictly speaking only ${}^{(s)}l^{j}(t_i)$ has dimensions of angular momentum, whereas ${}^{(s)}l_{j}$ is dimensionless, so that $l_s$ has units of $[l]^{-1}$. The total conserved $4$-momentum of the string, which we denote $\Pi_s$, is given by
\begin{eqnarray} \label{Pi}
\Pi_s = \sqrt{E_s^2-P_s^2-l_s^2}.
\end{eqnarray}
Finally we note that, in general, there may be many conserved charges but that, for a string, only two are independent.  The string configuration is ultimately invariant only under translations of the WS coordinates, $\tau \rightarrow \tau' = \tau + \delta\tau$, $\sigma \rightarrow \sigma' = \sigma + \delta\sigma$, and all other isometries are determined by these translations via the embedding. Therefore, only one independent conserved momentum exists in addition to the Hamiltonian.
%
\subsection{Long, straight strings with uniform geodesic windings} \label{Long, straight strings with uniform geodesic windings} 
Arguably, the simplest wound-string model in any extra-dimensional geometry is of a long, straight string in the large dimensions with geodesic windings in the compact manifold which are distributed uniformly along its length. The general embedding for a long string in the KS geometry with geodesic, though not necessarily uniform, windings in the $S^3$ is
\begin{eqnarray} \label{embedding1}
X^I(t,\sigma)=\left(t,x(t,\sigma)=0,y(t,\sigma)=0,z(t,\sigma)=(2\pi)^{-1}\Delta_s \sigma,0,0,0,\psi(t,\sigma)=0,\chi(t,\sigma)=0,\phi(t,\sigma)\right),
\end{eqnarray}
where 
\begin{eqnarray} \label{winding_condition}
\frac{1}{2\pi}\int_{0}^{2\pi}d\sigma \phi'(t,\sigma) = n_{\phi}, 
\end{eqnarray}
and $n_{\phi} \in \mathbb{Z}$ is the number of windings. Here, $a\Delta_s$ is the proper length of the string-section we are considering in the warped Minkowski directions and we have identified the WS time-coordinate $\tau$ with the proper time $t$ in the rest frame of the string centre-of-mass (CoM). Although, in reality, the string length will grow in proportion to the horizon distance, $d_H \propto at$, we consider a section of constant length.  In this case, the action becomes
\begin{eqnarray} \label{NG1}
S = -\mathcal{T}\int dt d\sigma \sqrt{a^2((2\pi)^{-2}a^2\Delta_s^2+R^2\phi'^2)-(2\pi)^{-2}a^2\Delta_s^2R^2\dot{\phi}^2}
\end{eqnarray}
and the constants of motion are
\begin{eqnarray} \label{P_01}
{}^{(s)}P_{0} &=& a^2{}^{(s)}P^{0} 
\nonumber\\
&=& \mathcal{T}\int_{0}^{2\pi}d\sigma \frac{a^2((2\pi)^{-2}a^2\Delta_s^2+R^2\phi'^2)}{\sqrt{a^2((2\pi)^{-2}a^2\Delta_s^2+R^2\phi'^2)-(2\pi)^{-2}a^2\Delta_s^2R^2\dot{\phi}^2}},
\end{eqnarray}
\begin{eqnarray} \label{P_z1*}
{}^{(s)}P_{z} &=& -a^2{}^{(s)}P^{z} 
\nonumber\\
&=& \mathcal{T}\int_{0}^{2\pi}d\sigma \frac{(2\pi)^{-1}a^2\Delta_s R^2\dot{\phi} \phi'}{\sqrt{a^2((2\pi)^{-2}a^2\Delta_s^2+R^2\phi'^2)-(2\pi)^{-2}a^2\Delta_s^2R^2\dot{\phi}^2}},
\end{eqnarray}
\begin{eqnarray} \label{l_phi1}
{}^{(s)}l_{\phi} &=& -R^2{}^{(s)}l^{\phi} 
\nonumber\\
&=& \mathcal{T}\int_{0}^{2\pi}d\sigma \frac{(2\pi)^{-2}a^2\Delta_s^2 R^2\dot{\phi}}{\sqrt{a^2((2\pi)^{-2}a^2\Delta_s^2+R^2\phi'^2)-(2\pi)^{-2}a^2\Delta_s^2R^2\dot{\phi}^2}}.
\end{eqnarray}
We now specify the ansatz for $\phi(t,\sigma)$ to represent uniform windings moving with constant velocity, $v_{\phi}=\dot{\phi}R<1$, in the compact space,
\begin{eqnarray}  \label{phi_ansatz1}
\phi(t,\sigma) = n_{\phi}\sigma + \frac{v_{\phi}t}{R},
\end{eqnarray}
so that 
\begin{eqnarray} \label{P_01A}
{}^{(s)}P_{0} = 2\pi \mathcal{T}\frac{a^2((2\pi)^{-2}a^2\Delta_s^2+R^2n_{\phi}^2)}{\sqrt{a^2((2\pi)^{-2}a^2\Delta_s^2+R^2n_{\phi}^2)-(2\pi)^{-2}a^2\Delta_s^2v_{\phi}^2}},
\end{eqnarray}
\begin{eqnarray} \label{P_z1A}
{}^{(s)}P_{z} &=& n_{\phi}{}^{(s)}l_{\phi}/((2\pi)^{-1}\Delta_s)
\nonumber\\
&=& 2\pi\mathcal{T} \frac{(2\pi)^{-1}a^2\Delta_s R v_{\phi} n_{\phi}}{\sqrt{a^2((2\pi)^{-2}a^2\Delta_s^2+R^2n_{\phi}^2)-(2\pi)^{-2}a^2\Delta_s^2v_{\phi}^2}}.
\end{eqnarray}
If the windings are produced dynamically by the motion of the string around the internal manifold, rather than being inherent as an initial condition from the moment of string formation, we expect $n_{\phi}$ to be given by
\begin{eqnarray} \label{n_{phi}1}
n_{\phi} = \frac{v_{\phi}\Delta_s}{2\pi R},
\end{eqnarray}
and imposing (\ref{n_{phi}1}) on (\ref{P_01A})-(\ref{P_z1A}) yields
\begin{eqnarray} \label{P_0_P^01}
{}^{(s)}P_{0} = \mathcal{T} a^2 \Delta_s \left(1 + \frac{(2\pi)^{2}R^2}{d_{\phi}^2}\right), \ \ \ {}^{(s)}P^{0} = \mathcal{T} \Delta_s \left(1 + \frac{(2\pi)^{2}R^2}{d_{\phi}^2}\right),
\end{eqnarray}
\begin{eqnarray} \label{P_z_P^z1} 
{}^{(s)}P_{z} = \mathcal{T} a n_{\phi} \times \frac{(2\pi)^{2}R^2}{d_{\phi}}, \ \ \ {}^{(s)}P^{z} = - \mathcal{T} \frac{n_{\phi}}{a} \times \frac{(2\pi)^{2} R^2}{d_{\phi}},
\end{eqnarray}
\begin{eqnarray} \label{l_{phi}_l^{phi}1}
{}^{(s)}l_{\phi} = 2\pi \mathcal{T} n_{\phi} R^2, \ \ \ {}^{(s)}l^{\phi} = -2\pi \mathcal{T} n_{\phi} , 
\end{eqnarray}
where we have defined $d_{\phi}$ as the distance between windings or, strictly speaking, as the distance between identical points on neighbouring windings,
\begin{eqnarray} \label{d_{phi}1}
d_{\phi} = \frac{a\Delta_s}{n_{\phi}}.
\end{eqnarray}
Equivalently, we may describe the state of the string using the total energy and momenta, $E_s$, $P_s$ and $l_s$, defined in (\ref{E,P,l});
\begin{eqnarray} \label{E_s1}
E_s = \mathcal{T} a\Delta_s \left(1 + \frac{(2\pi)^2R^2}{d_{\phi}^2}\right),
\end{eqnarray}
\begin{eqnarray} \label{P1}
P_s = (2\pi)^2 \mathcal{T} n_{\phi} \frac{R^2}{d_{\phi}},
\end{eqnarray}
\begin{eqnarray} \label{l1}
l_s = 2\pi \mathcal{T} n_{\phi} R,
\end{eqnarray}
or via the total $4$-momentum, $\Pi_s$, defined in (\ref{Pi});
\begin{eqnarray} \label{Pi1}
\Pi_s = \mathcal{T} a \Delta_s \sqrt{1 + \frac{(2\pi)^2R^2}{d_{\phi}^2}}.
\end{eqnarray}
\subsection{String loops with uniform, geodesic windings} \label{String loops with uniform, geodesic windings} 
We may generalize the model presented in Sect. \ref{Long, straight strings with uniform geodesic windings}  to circular string loops by first switching coordinate systems and allowing 
\begin{eqnarray} \label{coord_transf}
\left\{x(t,\sigma),y(t,\sigma),z(t,\sigma)\right\} \rightarrow \left\{x(t,\sigma),\rho(t),\sigma\right\},
\end{eqnarray}
in the ansatz (\ref{embedding1}), giving
\begin{eqnarray} \label{embedding2}
X^I(t,\sigma)=\left(t,x(t,\sigma)=0,\rho_s(t),\sigma,0,0,0,\psi(t,\sigma)=0,\chi(t,\sigma)=0,\phi(t,\sigma)\right),
\end{eqnarray}
before specifying the form of $\phi(t,\sigma)$, so that
\begin{eqnarray}  \label{phi_ansatz2}
\phi(t,\sigma) = n_{\phi}\sigma + \varphi(t).
\end{eqnarray}
Eq. (\ref{phi_ansatz2}) is the generalization of (\ref{phi_ansatz1}) and allows the velocity of the string in the compact space, which we will label 
\begin{eqnarray} \label{nu_{phi}(t)}
\nu_{\phi}(t)=\dot{\phi}(t)R, 
\end{eqnarray}
to vary in time in response to the time-evolution of the loop radius, $\rho_s(t)$, which is necessary to maintain conservation of energy and momentum. The constants of motion now become
\begin{eqnarray} \label{P_02}
{}^{(s)}P_{0} &=& a^2{}^{(s)}P^{0} 
\nonumber\\
&=& \mathcal{T}\int_{0}^{2\pi}d\sigma \frac{a^2(a^2\rho_s^2+R^2\phi'^2)}{\sqrt{a^2(1-\dot{\rho}_s^2)(a^2\rho_s^2+R^2\phi'^2)-a^2\rho_s^2R^2\dot{\phi}^2}},
\end{eqnarray}
\begin{eqnarray} \label{P_z1}
{}^{(s)}l_{\sigma} &=& -a^2\rho_s^2{}^{(s)}l^{\sigma}(t) 
\nonumber\\
&=& \mathcal{T}\int_{0}^{2\pi}d\sigma \frac{a^2\rho_s^2 R^2\dot{\phi} \phi'}{\sqrt{a^2(1-\dot{\rho}_s^2)(a^2\rho_s^2+R^2\phi'^2)-a^2\rho_s^2R^2\dot{\phi}^2}},
\end{eqnarray}
\begin{eqnarray} \label{l_phi1}
{}^{(s)}l_{\phi} &=& -R^2{}^{(s)}l^{\phi} 
\nonumber\\
&=& \mathcal{T}\int_{0}^{2\pi}d\sigma \frac{a^2\rho_s^2 R^2\dot{\phi}}{\sqrt{a^2(1-\dot{\rho}_s^2)(a^2\rho_s^2+R^2\phi'^2)-a^2\rho_s^2R^2\dot{\phi}^2}},
\end{eqnarray}
where $\phi'(t,\sigma) = n_{\phi}$, as before. We may create a model of a loop chopping-off from a long-string network by first assuming that the ansatz (\ref{embedding1}) holds, at least approximately, up until the time of loop formation, $t_i$. We then perform the coordinate transformation (\ref{coord_transf}) and identify
\begin{eqnarray} \label{identify}
\rho_s(t_i) = (2\pi)^{-1}\Delta_s, \ \ \ \dot{\phi}(t_i)R=\nu_{\phi}(t_i) = v_{\phi}, \ \ \ \frac{{}^{(s)}l_{\sigma}}{2\pi \rho_s(t_i)} = {}^{(s)}P_{z}
\end{eqnarray}
together with imposing the initial condition
\begin{eqnarray} \label{bc*}
\dot{\rho}_s(t_i) = 0.
\end{eqnarray}
In the limit $a\rho_s(t_i)>>R$, this seems reasonable, since we may consider the string to be approximately linear, so that $z(t_i,\sigma) =(2\pi)^{-1}\Delta_s\sigma \approx \rho_s(t_i)\sigma$. However, we must switch coordinate systems and abandon the linear approximation in order to treat the time-evolution of $\rho_s(t)$ for $t>t_i$. This is also necessary in order to account for the fact that the linear momentum, ${}^{(s)}P_z$, is no longer conserved and the ``angular momentum" ${}^{(s)}l_{\sigma}$ takes its place. The analogue of (\ref{n_{phi}1}) is 
\begin{eqnarray} \label{n_{phi}1A} 
n_{\phi} = \frac{\nu_{\phi}(t_i)\rho_s(t_i)}{R} 
\end{eqnarray}
and the boundary conditions (\ref{identify})-(\ref{n_{phi}1A}), yield  
\begin{eqnarray} \label{P_0_P^02}
{}^{(s)}P_{0} = 2\pi\mathcal{T} a^2 \rho_s(t_i) \left(1 + \frac{(2\pi)^2R^2}{d_{\phi}^2(t_i)}\right), \ \ \ {}^{(s)}P^{0} = 2\pi\mathcal{T} \rho_s(t_i) \left(1 + \frac{(2\pi)^2R^2}{d_{\phi}^2(t_i)}\right),
\end{eqnarray}
\begin{eqnarray} \label{l_{sigma}_l^{sigma}1}
{}^{(s)}l_{\sigma} = (2\pi)^2 \mathcal{T} n_{\phi}^2 R^2, \ \ \ {}^{(s)}l^{\sigma}(t) = -(2\pi)^2 \mathcal{T} \frac{n_{\phi}^2 R^2}{a^2\rho_s^2(t)},
\end{eqnarray}
together with (\ref{l_{phi}_l^{phi}1}), where we have defined
\begin{eqnarray} \label{d_{phi}2}
d_{\phi}(t) = \frac{2\pi a\rho_s(t)}{\sqrt{1-\dot{\rho}_s^2(t)}n_{\phi}},
\end{eqnarray}
which is the generalization of (\ref{d_{phi}1}). The expressions in (\ref{P_0_P^02}) are equivalent to those in (\ref{P_0_P^01}) under the identifications (\ref{identify}), together with (\ref{bc*}), and the value of ${}^{(s)}l^{\sigma}(t_i)$, according to (\ref{l_{sigma}_l^{sigma}1}), is given via $2\pi \rho_s(t_i) {}^{(s)}l^{\sigma}(t_i)={}^{(s)}P^{z}$. This relation ensures that ${}^{(s)}l_{\sigma}{}^{(s)}l^{\sigma}(t_i)={}^{(s)}P_z{}^{(s)}P^z$. Likewise, $E_s$ and $l_s$ are given by
\begin{eqnarray} \label{E_s2}
E_s = 2\pi \mathcal{T} a\rho_s(t_i) \left(1 + \frac{(2\pi)^2R^2}{d_{\phi}^2(t_i)}\right),
\end{eqnarray}
\begin{eqnarray} \label{l2}
l_s = 2\pi \mathcal{T}n_{\phi} R \sqrt{1+\frac{(2\pi)^2R^2}{d_{\phi}^2(t_i)}},
\end{eqnarray}
and yield the following expression for $\Pi_s$,
\begin{eqnarray} \label{Pi2}
\Pi_s =  2\pi \mathcal{T}a\rho_s(t_i) \sqrt{1 + \frac{(2\pi)^2R^2}{d_{\phi}^2(t_i)}}, 
\end{eqnarray}
which is equivalent to (\ref{Pi1}) under  (\ref{identify})-(\ref{n_{phi}1A}) and (\ref{d_{phi}2}). The difference between the two regimes therefore lies, not in the value of the total $4$-momentum, which must be conserved at the epoch of loop formation according to the matching conditions (\ref{identify})-(\ref{bc*}), but in the time evolution of $\rho_s(t)$ and $\varphi(t)$, whose EOM we can now derive using the constants determined above. Combining (\ref{P_02}) and (\ref{P_0_P^02}) gives the EOM for $\rho_s(t)$;
\begin{eqnarray} \label{loop_EOM1}
\dot{\rho}_s^2-1 +\left(\frac{a^2\rho_s^2(t_i)}{a^2\rho_s^2(t_i)+R^2n_{\phi}^2}\right)^2 \left(\frac{\rho_s}{\rho_s(t_i)}+\frac{R^2n_{\phi}^2}{a^2\rho_s(t_i)\rho_s}\right)^2 = 0.
\end{eqnarray}
The solution of Eq.  (\ref{loop_EOM1}) may be written in a convenient form by first defining the parameter $\omega_{\phi}^2(t)$, which represents  the (time-dependent), fraction of the total string length in the compact space 
\begin{eqnarray} \label{omega_{phi}(t)}
\omega_{\phi}^2(t) = \frac{R^2n_{\phi}^2}{a^2\rho_s^2(t)+R^2n_{\phi}^2},
\end{eqnarray}
and its counterpart, $\omega_{s\sigma}^2(t)=1-\omega_{\phi}^2(t)$, which represents the remaining fraction in the infinite directions,
\begin{eqnarray} \label{omega_{sigma}(t)}
\omega_{s\sigma}^2(t) = \frac{a^2\rho_s^2(t)}{a^2\rho_s^2(t)+R^2n_{\phi}^2}.
\end{eqnarray}
In terms of  $\omega_{s\sigma}^2(t_i)$, the time-evolution of $\rho_s(t)$ is \cite{Cycloops, Thesis}
\begin{eqnarray}  \label{loop_EOM_soln1}
\rho_s(t) = \rho_s(t_i)\sqrt{1 + \left(\frac{1-2\omega_{s\sigma}^2(t_i)}{\omega_{s\sigma}^4(t_i)}\right)\sin^2\left(\frac{\omega_{s\sigma}^2(t_i)}{\rho_s(t_i)}(t-t_i)\right)},
\end{eqnarray}
so that the string oscillates between the following critical values of $\rho_s(t)$, $d_{\phi}(t)$ and $\omega_{s\sigma}^2(t)$, with time period $T=(p+1/2)\pi \rho_s(t_i)/\omega_{s\sigma}^2(t_i)$, $p\in \mathbb{Z}$,
\begin{eqnarray} \label{crit_values}
\rho_{s(c1)} &=& \rho_s(t_i), \ \ \ \rho_{s(c2)}=\frac{1-\omega_{s\sigma}^2(t_i)}{\omega_{s\sigma}^2(t_i)} \rho_s(t_i),
\nonumber\\
d_{\phi(c1)} &=& 2\pi\frac{\omega_{s\sigma}(t_i)}{\sqrt{1-\omega_{s\sigma}^2(t_i)}}R, \ \ \ d_{\phi(c2)} = 2\pi\frac{\sqrt{1-\omega_{s\sigma}^2(t_i)}}{\omega_{s\sigma}(t_i)}R,
\nonumber\\
\omega_{s\sigma(c1)}^2 &=& \omega_{s\sigma}^2(t_i), \ \ \ \omega_{s\sigma(c2)}^2 = 1- \omega_{s\sigma}^2(t_i),
\end{eqnarray}
where we have used the alternative formula for $n_{\phi}$
\begin{eqnarray} \label{n_phi_approx1}
n_{\phi} = \frac{\omega_{\phi}(t_i)\rho_s(t_i)}{R},
\end{eqnarray}
and set $\omega_{s\sigma}^2(t_i) = a^2$, to obtain the values of $d_{\phi(c1)}$ and $d_{\phi(c2)}$. We will return to these definitions in Sect. \ref{Specific, but natural, models for winding formation}, when we will see that, in general, the formula for $n_{\phi}$ may be written in terms of the ratio of the velocities in the compact and large dimensions or, equivalently, in terms of the fractions of the total string length in each, which may be written as functions of the warp factor.
\\ \indent
For $\omega_{s\sigma}^2(t_i)<=>1/2$, we have that $\rho_{s(c2)}>=<\rho_s(t_i)$ and $d_{\phi(c2)}>=<2\pi R$. Interestingly, in the critical case, the string is prevented from oscillating under the action of its own tension by the conservation of angular momentum in the compact space, though such configuration is clearly unstable to perturbations, since even a slight deviation from the critical value will cause the loop to fall into one of the two other dynamical regimes \cite{Cycloops,Thesis}. 
\\ \indent
Combining (\ref{P_z1}) with (\ref{l_{sigma}_l^{sigma}1}) gives the evolution of $\dot{\phi}$ in terms of $\rho_s(t)$ and $\dot{\rho}_s(t)$, which shows that $\dot{\phi}$ and $\rho_s(t)$ evolve so as to keep the number of windings, $n_{\phi}$, fixed
\begin{eqnarray} \label{n_{phi}2}
n_{\phi} =  \frac{\rho_s(t)}{\sqrt{1-\dot{\rho}_s^2(t)}}\dot{\phi}(t)=\frac{\rho_s(t)}{\sqrt{1-\dot{\rho}_s^2(t)}} \frac{\nu_{\phi}(t)}{R}.
\end{eqnarray}
This is to be expected. Although, physically, a change in $n_{\phi}$ is of course possible, since the windings are not topologically stabilized, we are not able to account for this possibility using an embedding of the form (\ref{embedding2}). However, explicit conditions for dynamical stabilization were derived in \cite{BlancoPillado} and are reviewed briefly in Sect. \ref{Stability analysis}. Using the definition (\ref{d_{phi}2}), Eq. (\ref{n_{phi}2}) may be rewritten as
\begin{eqnarray} \label{reln1}
\dot{\phi}(t)d_{\phi}(t) = 2\pi a. 
\end{eqnarray}
Comparing (\ref{n_{phi}2}) with (\ref{n_phi_approx1}) also implies that
\begin{eqnarray} 
\nu(t_i) = \omega_{\phi}(t_i) = \sqrt{1-\omega_{s\sigma}^2(t_i)}.
\end{eqnarray}
Finally, we note that the EOM (\ref{loop_EOM1}) and the relation (\ref{n_{phi}2}) may also be obtained simply by taking the identification (\ref{identify}) and allowing all quantities that can become time-dependent, to become time-dependent for $t>t_i$ (including Lorentz factors where appropriate), under the condition that the constants of motion (and $n_{\phi}$) remain unchanged. That is, by allowing
\begin{eqnarray} \label{ident}
\Delta_s &\rightarrow& \frac{2\pi \rho_s(t_i)}{\sqrt{1-\dot{\rho}_s^2(t_i)}} \rightarrow \frac{ 2\pi \rho_s(t)}{\sqrt{1-\dot{\rho}_s^2(t)}}, 
\nonumber\\
d_{\phi} &\rightarrow& d_{\phi}(t_i) = \frac{ 2\pi a \rho_s(t_i)}{\sqrt{1-\dot{\rho}_s^2(t_i)}n_{\phi}} \rightarrow \sqrt{1-\dot{\rho}_s^2(t)}d_{\phi}(t) =  \frac{ 2\pi a\rho_s(t)}{n_{\phi}},
\nonumber\\
v_{\phi} &\rightarrow& \nu_{\phi}(t_i) = \dot{\phi}(t_i)R \rightarrow \nu_{\phi}(t) = \dot{\phi}(t)R,
\end{eqnarray}
and imposing (\ref{bc*}) at $t=t_i$, we obtain (\ref{n_{phi}2}) as the generalization of (\ref{n_{phi}1}), together with
\begin{eqnarray} \label{E_s3}
E_s &=&  2\pi \mathcal{T} \frac{a\rho_s(t)}{\sqrt{1-\dot{\rho}_s^2(t)}} \left(1 + \frac{R^2n_{\phi}^2}{a^2\rho_{s}^2(t)}\right) 
=  2\pi \mathcal{T} a\rho_s(t_i) \left(1 + \frac{R^2n_{\phi}^2}{a^2\rho_{s}^2(t_i)}\right)
\end{eqnarray}
as the generalization of (\ref{E_s1}), which also gives rise to (\ref{loop_EOM1}). Eqs. (\ref{l_{phi}_l^{phi}1}), (\ref{l_{sigma}_l^{sigma}1}) and (\ref{l2}) follow directly. This is an important point since it implies that, even if we were unable to derive the EOM for a string loop using the ansatz (\ref{embedding2}), we would still be able to determine the macroscopic behavior of the loop by using the straight-string ansatz as an approximation up to $t=t_i$, before imposing the matching conditions (\ref{identify})-(\ref{bc*}) and allowing all quantities which are able to vary in time to do so, under the condition that $\Pi_s$ and $n_{\phi}$ are conserved. We will make use of this method in Sect. \ref{Transition to loops in the field theory}, when an exact ansatz which satisfies the fundamental abelian-Higgs EOM and which corresponds to the field configuration for a looped topological defect string with twisted magnetic flux lines is difficult to obtain directly. 
\section{Recap of the abelian-Higgs Model} \label{Review of the abelian-Higgs Model} 
In this section we briefly review the essential points of the abelian-Higgs model, with special emphasis on the the nature of the conserved quantities in cylindrical polar coordinates $\left\{t,r,\theta,z\right\}$, which we will use in Sect. \ref{Field configuration for a long, straight string with twisted flux-lines: Model 1}  to describe a long, straight, topological defect string, and on the matching conditions which are the analogues of (\ref{identify})-(\ref{bc*}). The basic action for the abelian-Higgs model, with external charged matter, $j^{\mu}$, is
\begin{eqnarray} \label{eq:action}
S = \int d^4x \sqrt{-g} \left\{D_{\mu}\phi \overline{D}^{\mu}\overline{\phi} - \frac{1}{4}F_{\mu\nu}F^{\mu\nu} - V(|\phi|) - eA_{\mu}j^{\mu}\right\},
\end{eqnarray}
where $\mu, \nu \in \left\{0,1,2,3\right\}$ and $V(|\phi|)$ is the potential term, which is determined by the symmetry-breaking energy-scale, $\eta$, and the dimensionless scalar coupling, $\lambda$;
\begin{eqnarray} \label{potential}
V(|\phi|) = \frac{\lambda}{2}(|\phi|^2-\eta^2)^2.
\end{eqnarray}
In this paper, the gauge covariant derivative, $D_{\mu}$, and its conjugate, $\overline{D}_{\mu}$, are defined according to
\begin{eqnarray} \label{cov_deriv}
D_{\mu} = \partial_{\mu} - ieA_{\mu}, 
\nonumber\\
\overline{D}_{\mu} = \partial_{\mu} + ieA_{\mu}.
\end{eqnarray}
The electromagnetic field tensor, $F_{\mu\nu}$, is defined in the usual way,
\begin{eqnarray}  \label{EM_tens}
F_{\mu\nu} = \partial_{\mu}A_{\nu} - \partial_{\nu}A_{\mu},
\end{eqnarray}
and we use the metric convention $(+---)$, as in Sect. \ref{Wound-strings in the KS geometry}, throughout, so that the action is invariant under local transformations of the form
\begin{eqnarray} \label{gauge_transf}
\phi(x) \rightarrow \phi'(x) = \phi(x)e^{i\Lambda(x)}, \ \ \ A_{\mu}(x) \rightarrow A'_{\mu}(x) = A_{\mu}(x) + \frac{1}{e}\partial_{\mu}\Lambda(x).
\end{eqnarray}
The covariant form of the scalar, scalar-conjugate and vector EOM are, respectively,
\begin{eqnarray} \label{cov_EOMs}
\frac{1}{\sqrt{-g}}D_{\mu}\left(\sqrt{-g}D^{\mu}\phi\right) + \frac{\lambda}{2}\phi\left(|\phi|^2-\eta^2\right) = 0,
\nonumber\\
\frac{1}{\sqrt{-g}}\overline{D}_{\mu}\left(\sqrt{-g}\overline{D}^{\mu}\overline{\phi}\right) + \frac{\lambda}{2}\overline{\phi}\left(|\phi|^2-\eta^2\right) = 0,
\end{eqnarray}
\begin{eqnarray} \label{eq:EOM_vector_Abelian-Higgs_L}  
\frac{1}{\sqrt{-g}}\partial_{\mu}\left(\sqrt{-g}F^{\mu\nu}\right) + ie\left(\overline{\phi}D^{\nu}\phi-\phi\overline{D}^{\nu}\overline{\phi}\right) - ej^{\nu}= 0,
\end{eqnarray}
and the total current is given by
\begin{eqnarray} \label{J}
J^{\nu}(x) = \mathcal{J}^{\nu}(x) + j^{\nu}(x) = \frac{1}{\sqrt{-g}}\partial_{\mu}(\sqrt{-g}F^{\mu\nu})e^{-1},
\end{eqnarray}
where $\mathcal{J}^{\nu}(x)$ is the $U(1)$ current,
\begin{eqnarray} \label{mathcal{J}}
\mathcal{J}^{\nu} = -i(\overline{\phi}D^{\mu}\phi-\phi\overline{D}^{\mu}\overline{\phi}),
\end{eqnarray}
and total conserved charge is quantized in terms of $e$;
\begin{eqnarray} \label{Q}
Q = e\int d^3x\sqrt{-g}J^{0}. 
\end{eqnarray}
The energy-momentum tensor, $T^{\mu}{}_{\nu}$, is
\begin{eqnarray} \label{eq:T_mu_nu_2}
T^{\mu}{}_{\nu}  = \frac{-2}{\sqrt{-g}}g_{\alpha\nu}\frac{\delta S}{\delta g_{\mu\alpha}}
= D^{\mu}\phi\overline{D}_{\nu}\overline{\phi} + D_{\nu}\phi\overline{D}^{\mu}\overline{\phi}
- F^{\mu\alpha}F_{\nu}{}_{\alpha} - e(A^{\mu}j_{\nu} + A_{\nu}j^{\mu}) - \delta^{\mu}{}_{\nu}\mathcal{L},
\end{eqnarray}
which gives rise to the conserved charges
\begin{eqnarray} \label{cons}
\Pi_{\mu} = \int  d^3x\sqrt{-g} T^{0}{}_{\mu}.
\end{eqnarray}
Corresponding quantities with contravariant indices, $\Pi^{\mu}$, are defined (as in Sect. \ref{Wound-strings in the KS geometry}), via 
 \begin{eqnarray} \label{notcons}
\Pi^{\mu} = \int d^3x\sqrt{-g} T^{0\mu},
\end{eqnarray}
and, since these represent components with units of energy (when $\mu=0$), and either linear or angular momentum (when $\mu \neq 0$), depending upon the coordinate system, we again use the notation $\Pi^{\mu}$ only when the nature of the conserved charge remains unspecified and instead use $P^{0}$, $P^i$ or $l^j$ (and their covariant counterparts), to denote quantities related to time-translational invariance, rectilinear space-translational invariance and rotational invariance, as appropriate. Yet again, we will find that contravariant quantities with dimensions $[l]^{-2}$ are not necessarily time-independent, in which case $l^{j}(t_i)$ represents the conserved angular momentum. By analogy with (\ref{E,P,l})-(\ref{Pi}), we define the quantities $E_{|n|}$, $P_{|n|}$, $l_{|n|}$ and $\Pi_{|n|}$, via
\begin{eqnarray} \label{E,P,l2}
E_{|n|} = \sqrt{P_0P^{0}}, \ \ \ P_{|n|}=\sqrt{-P_{i}P^{i}}, \ \ \ l_{|n|}=\sqrt{-l_{j}l^{j}(t_i)},
\end{eqnarray}
\begin{eqnarray} \label{Pi2}
\Pi_{|n|} = \sqrt{E_{|n|}^2-P_{|n|}^2-l_{|n|}^2},
\end{eqnarray}
though since, in this model, we are not working in warped space, we have that $\Pi^{\mu}=\pm \Pi_{\mu}$, except when $\mu$ corresponds to an angular coordinate, and we do not have to worry about powers of the warp factor coming from the metric. In this case the subscript $|n|$ refers to the fact that we will be investigating vortex-configurations and, in general, the quantities above will depend on the value of the topological winding number of the defect string. The Hamiltonian is therefore given by
\begin{eqnarray} \label{eq:Hamiltonian}
E_{|n|} &=& P_{0} = \int T^{0}{}_{0} d^{3}x\sqrt{-g}
\end{eqnarray}
which, in cylindrical polars (with $c=1$ as before), may be written as
\begin{eqnarray} \label{eq:H_in_cp_coords}
E_{|n|} &=& \int rdrd\theta dz\left\{\left|\frac{\partial\phi}{\partial t}-ieA_{0}\phi\right|^2 + \left|\frac{\partial\phi}{\partial r}-ieA_{r}\phi\right|^2
+ \frac{1}{r^2}\left|\frac{\partial\phi}{\partial\theta}-ieA_{\theta}\phi\right|^2 + \left|\frac{\partial\phi}{\partial z}-ieA_{z}\phi\right|^2\right\}
\nonumber\\
&+& \int rdrd\theta dz\left\{\frac{1}{2}(\vec{E}^2 + \vec{B}^2) + \frac{\lambda}{2}(|\phi|^2-\eta^2)^2 - e(A_0j_0 + \vec{A}.\vec{j})\right\},
\end{eqnarray}
where we have defined
\begin{eqnarray}
\vec{E}^2 = \left(\frac{\partial A_{r}}{\partial t}-\frac{\partial A_{0}}{\partial r}\right)^2
+ \frac{1}{r^2}\left(\frac{\partial A_{\theta}}{\partial t}-\frac{\partial A_{0}}{\partial \theta}\right)^2
+ \left(\frac{\partial A_{z}}{\partial t}-\frac{\partial A_{0}}{\partial z}\right)^2,
\end{eqnarray}
\begin{eqnarray}
\vec{B}^2 = \frac{1}{r^2}\left(\frac{\partial A_{\theta}}{\partial z}-\frac{\partial A_{z}}{\partial \theta}\right)^2
+\left(\frac{\partial A_{z}}{\partial r}-\frac{\partial A_{r}}{\partial z}\right)^2
+ \frac{1}{r^2}\left(\frac{\partial A_{r}}{\partial\theta}-\frac{\partial A_{\theta}}{\partial r}\right)^2,
\end{eqnarray}
\begin{eqnarray}
 \vec{A}.\vec{j} = A_r j_r + \frac{1}{r^2}A_{\theta}j_{\theta} + A_z j_z,
\end{eqnarray}
and the covariant components of the conserved momentum are
\begin{eqnarray} \label{eq:P^z}
P_{z} &=& \int T^{0}{}_{z} rdrd\theta dz
\nonumber\\ 
 T^{0}{}_{z} &=& \left(\frac{\partial\phi}{\partial t}-ieA_{0}\phi\right)\left(\frac{\partial\overline{\phi}}{\partial z}+ieA_{z}\overline{\phi} \right) + e(A_{0}j_{z} + A_{z}j_{0})
\nonumber\\
&+& \frac{1}{2}\left[\left(\frac{\partial A_{r}}{\partial t}-\frac{\partial A_{0}}{\partial r}\right)\left(\frac{\partial A_{r}}{\partial z}-\frac{\partial A_{z}}{\partial r}\right)
+ \frac{1}{r^2}\left(\frac{\partial A_{\theta}}{\partial t}-\frac{\partial A_{0}}{\partial \theta}\right)\left(\frac{\partial A_{\theta}}{\partial z}-\frac{\partial A_{z}}{\partial \theta}\right)\right],
\end{eqnarray}
\begin{eqnarray} \label{eq:P^theta}
l_{\theta} &=& \int T^{0}{}_{\theta} rdrd\theta dz
\nonumber\\ 
T^{0}{}_{\theta} &=& \left(\frac{\partial\phi}{\partial t}-ieA_{0}\phi\right)\left(\frac{\partial\overline{\phi}}{\partial \theta}+ieA_{\theta}\overline{\phi}\right) + e(A_{0}j_{\theta} + A_{\theta}j_{0})
\nonumber\\
&+& \frac{1}{2}\left[\left(\frac{\partial A_{z}}{\partial t}-\frac{\partial A_{0}}{\partial z}\right)\left(\frac{\partial A_{z}}{\partial \theta}-\frac{\partial A_{\theta}}{\partial z}\right)
+ \left(\frac{\partial A_{r}}{\partial t}-\frac{\partial A_{0}}{\partial r}\right)\left(\frac{\partial A_{r}}{\partial \theta}-\frac{\partial A_{\theta}}{\partial r}\right)\right].
\end{eqnarray}
As stated above, the contravariant component $l^{\theta}$, evaluated using $T^{0}{}_{\theta}=-r^2T^{0\theta}$, has dimensions $[l]^{-2}$ and may be interpreted as an angular momentum, whereas $P^{z}=-P_{z}$ has dimensions $[l]^{-1}$ and may be interpreted as a linear momentum. In addition, if the $\theta$-coordinate represents the angle around the central axis of a vortex-string, then both the dimensionless covariant component, $l_{\theta}$, and $l^{\theta}$ are directly conserved. However, if we make the transition from a long, approximately straight string section, whose total length, $\Delta$, remains fixed, to a string loop whose radius evolves in time, $\rho=\rho(t)$, using a transformation like that given in (\ref{coord_transf}), we may again identify $\rho(t_i)=(2\pi)^{-1}\Delta$ and $l_{\sigma}\rho^{-1}(t_i) = -l^{\sigma}(t_i)\rho(t_i)= -P^z$, but the quantity $l^{\sigma}(t)=-\rho^{-2}(t)l_{\sigma}= \rho(t_i)\rho^{-2}(t)P^z$ is, in general, time-dependent. 
\section{Field configuration for a long, straight string with twisted flux-lines} \label{Field configuration for a long, straight string with twisted flux-lines: Model 1} 
The ansatz we will use to describe strings with twisted lines of magnetic flux is of the general form
\begin{eqnarray} \label{eq:ansatz_phi} 
\phi(r,\theta,z,t) = \eta f(r) \exp \left[in\theta + in(k_{z}z + \omega_{z}t)\right],
\end{eqnarray}
\begin{eqnarray} \label{eq:ansatz_Ar*}
A_{r} = 0, \ \ \ A_{\theta}(r) = \frac{n}{e}a_{\theta}(r),
\end{eqnarray}
\begin{eqnarray}  \label{eq:ansatz_Az*}
A_{z} =\frac{n}{e}k_{z}a(r), \ \ \ A_{0} = \frac{n}{e}\omega_z a(r),
\end{eqnarray}
where $f(r)$, $a_{\theta}(r)$ and $a(r)$ are dimensionless functions and $k_z$, $\omega_z$ are constants with dimensions $[l]^{-1}$. Substituting (\ref{eq:ansatz_phi})-(\ref{eq:ansatz_Az*}) into (\ref{cov_EOMs})-(\ref{eq:EOM_vector_Abelian-Higgs_L}) gives
\begin{eqnarray} \label{eq:scalar_spec(1)G} 
\frac{d^2f}{dr^2} + \frac{1}{r}\frac{df}{dr} - \frac{n^2f}{r^2}(1-a_{\theta})^2 +  n^2f\left(1-a\right)^2(\omega_z^2-k_z^2) - \frac{1}{2}\frac{1}{r_s^2}f(f^2-1) = 0,
\end{eqnarray}
for the scalar EOM, plus
\begin{eqnarray} \label{eq:gauge_spec1(1)}
\frac{d^2a_{\theta}}{dr^2} - \frac{1}{r}\frac{da_{\theta}}{dr} + \frac{2}{r_{v}^2}f^2(1-a_{\theta}) + \frac{e^2}{n}j_{\theta}= 0,
\end{eqnarray}
for the $\theta$-component of the vector EOM and 
\begin{eqnarray} \label{eq:gauge_spec2(1)G}
\frac{d^2a}{dr^2} + \frac{1}{r}\frac{da}{dr} + \frac{2}{r_{v}^2}f^2(1-a) + \frac{e^2}{n}k_z^{-1}j_{z}= 0,
\end{eqnarray}
\begin{eqnarray} \label{eq:gauge_spec2(1)G*}
\frac{d^2a}{dr^2} + \frac{1}{r}\frac{da}{dr} + \frac{2}{r_{v}^2}f^2(1-a) + \frac{e^2}{n}\omega_z^{-1}j_{0}= 0,
\end{eqnarray}
for the $z$- and $t$-components, respectively. To obtain a unique equation for $a(r)$, we must therefore set
\begin{eqnarray} \label{j_zj_0}
j_z(r) &=& \frac{n}{e^2} k_z j(r),
\nonumber\\
j_0(r) &=& \frac{n}{e^2} \omega_z j(r),
\end{eqnarray}
so that, imposing the dispersion relation
\begin{eqnarray} \label{disp_reln}
\omega_z^2 = k_z^2, \ \ \  \omega_z = \pm k_z
\end{eqnarray}
and setting $j_{\theta}=0$ for simplicity, Eqs. (\ref{eq:scalar_spec(1)G})-(\ref{eq:gauge_spec1(1)}) reduce to the scalar EOM and the $\theta$-component of the vector EOM for the standard Nielsen-Olesen (NO) ansatz \cite{NO}; 
\begin{eqnarray} \label{NO_scalar} 
\frac{d^2f}{dr^2} + \frac{1}{r}\frac{df}{dr} - \frac{n^2f}{r^2}(1-a_{\theta})^2  - \frac{1}{2}\frac{1}{r_s^2}f(f^2-1) = 0,
\end{eqnarray}
\begin{eqnarray} \label{NO_vector}
\frac{d^2a_{\theta}}{dr^2} - \frac{1}{r}\frac{da_{\theta}}{dr} + \frac{2}{r_{v}^2}f^2(1-a_{\theta}) = 0,
\end{eqnarray}
while Eqs. (\ref{eq:gauge_spec2(1)G})-(\ref{eq:gauge_spec2(1)G*}) reduce to
\begin{eqnarray} \label{NO_vector+}
\frac{d^2a}{dr^2} + \frac{1}{r}\frac{da}{dr} + \frac{2}{r_{v}^2}f^2(1-a) + j= 0.
\end{eqnarray}
Eq. (\ref{disp_reln}) also implies
\begin{eqnarray} \label{disp_reln_impliesj}
j_0^2 = j_z^2, \ \ \  j_0 = \pm j_z,
\end{eqnarray}
and
\begin{eqnarray} \label{disp_reln_impliesA}
A_0^2 = A_z^2, \ \ \  A_0 = \pm A_z.
\end{eqnarray}
We now impose the following boundary conditions on the free functions $f(r)$, $a_{\theta}(r)$ and $a(r)$;
\begin{equation} \label{eq:f_bc}
f(r) = \left \lbrace
\begin{array}{rl}
0,& \ r=0 \\
1,& \ r \rightarrow \infty \ (r \gtrsim r_{s|n|}),
\end{array}\right.
\end{equation}
\begin{equation}\label{eq:alpha_bc}
a_{\theta}(r) = \left \lbrace
\begin{array}{rl}
0,& \ r=0 \\
1,& \ r \rightarrow \infty \ (r \gtrsim r_{v|n|}),
\end{array}\right.
\end{equation}
\begin{equation} \label{eq:alphap_bc}
a(r) = \left \lbrace
\begin{array}{rl}
0,& \  r=0 \\
1,& \  r \rightarrow \infty \ (r \gtrsim r_{v|n|}),
\end{array}\right.
\end{equation}
where $r_{s|n|}$, $r_{v|n|}$ are the scalar and vector core radii of a string with winding $\pm |n|$. Taking (\ref{eq:ansatz_phi})-(\ref{eq:ansatz_Az*}), we see that the ansatz for the $\theta$-component of the gauge field, $A_{\theta}(r)$, which gives rise to the $z$-component of the $B$-field, is identical to that in the NO ansatz, in which all other components of the EM tensor, besides $B^{z}(r)$, are zero. In addition, there now exist nonzero $A_{z}(r)$ and $A_{0}(r)$ terms which give rise to an additional magnetic field component, $B^{\theta}(r)$ and to a nonzero electric field, $E^{r}(r)$, respectively. 
\\ \indent
In the NO ansatz, a nonzero winding in the phase is required in order for the contribution to the energy density from the gradient term of the Hamiltonian to tend to zero as $r\rightarrow \infty$, and the  winding number must be equal to the number of units of magnetic flux. The flux-lines are directed purely along the $z$-direction and the winding in the phase is purely radial, lying in the plane perpendicular to $\vec{B}$. The lines of constant phase are therefore parallel to the lines of flux.
\\ \indent
In the gauge field ansatz above the flux-lines follow helical paths, determined by both $B^{z}(r)$ and $B^{\theta}(r)$, so we may expect that the lines of constant phase must do likewise, being determined by both a $\theta$-dependent term and a $z$-dependent term in the exponent which determines the phase of the scalar field. This, in turn, suggests the existence of an additional winding number along the length of the string. That is, following any line of fixed $(r,\theta)$ within the scalar vortex core, we expect the phase of $\phi$ to change by $\pm 2\pi|n|$ over some characteristic distance, $\lambda_z$, given by
\begin{eqnarray} \label{lambda_z}
k_z = \frac{2\pi}{\lambda_z}.
\end{eqnarray}
This does not contradict the fact that it is physically meaningless to ``rotate" a single (two-dimensional) vortex by some angle $\Delta\phi$. Such a transformation is simply a gauge mode, but there is nothing to prevent the phase at corresponding points in different vortex-slices within a single (three-dimensional) string by varying with respect to $z$. We will return to this point later in Sect. \ref{String energy and momentum}. 
\\ \indent 
Likewise, the nonzero $A_{0}(r)$ term in the gauge field also implies that we must include a term proportional to $t$ in the exponent of the phase factor for the scalar field. Together with the term proportional to $z$, this gives rise to helical lines of constant phase which rotate around the string central-axis. Physically, it may be possible to imagine such a field configuration forming if $\phi$ acquires some angular velocity as it rolls off the potential hill into the circle of degenerate minima during the phase-transition epoch, though, as mentioned in the introduction, a small amount of ``external" charged matter must be present within the string core in order for the energy and total charge to remain finite. This is demonstrated explicitly in our analysis of the EOM (\ref{NO_scalar})-(\ref{NO_vector+}) and their solutions. Firstly, let us assume that
\begin{eqnarray} \label{eq:rsn}
r_{s|n|}  \approx |n|^{\sigma}(\sqrt{\lambda}\eta)^{-1},
\end{eqnarray}
\begin{eqnarray} \label{eq:rvn}
r_{v|n|} \approx |n|^{\epsilon}\left(e \eta\right)^{-1},
\end{eqnarray}
where $\sigma \geq 0$, $\epsilon \geq 0$. We also use the simplified notation, $r_s$ and $r_v$, to signify the scalar and vector core radii of a string with winding $n=\pm 1$. It is common in the literature to assume $\sigma=0$ for NO strings, so that $r_{s|n|}  = r_s \approx (\sqrt{\lambda}\eta)^{-1}$
is independent of $|n|$. A value of $\epsilon = 1/2$ is then required in order to saturate the Bogomol'nyi-Prasad-Sommerfield (BPS) bound \cite{Bogomolnyi:1976_1}-\cite{de_Vega:1976_1}, 
\begin{eqnarray} \label{Bog}
{}^{(NO)}\mu_{|n|} \geq 2\pi \eta^2 |n|.
\end{eqnarray}
to stabilize the vortex. However, although the bound above holds for NO strings, it is not immediately clear whether it extends to strings described by the ansatz presented here, or whether values of $\sigma=0$, $\epsilon=1/2$ are required to ensure the stability.  For this reason, we take a more phenomenological approach and leave $\sigma$ and $\epsilon$ as free parameters for the time being. Since Eqs. (\ref{NO_scalar})-(\ref{NO_vector}) are simply the usual scalar and vector EOM for the NO ansatz, their solutions, to leading order and subject to the boundary conditions (\ref{eq:f_bc})-(\ref{eq:alpha_bc}), are well known and are given by;
\begin{eqnarray} \label{eq:f_lo}
f(r) \approx \left \lbrace
\begin{array}{rl}
(r/r_{s|n|})^{|n|},& \ 0 \leq r \lesssim r_{s|n|} \\
1,& \ r \rightarrow \infty, \ (r \gtrsim r_{s|n|}),
\end{array}\right.
\end{eqnarray}
\begin{eqnarray} \label{eq:atheta_lo}
a_{\theta}(r) \approx \left \lbrace
\begin{array}{rl}
(r/r_{v|n|})^2,& \ 0 \leq r \lesssim r_{v|n|} \\
1,& \ r \rightarrow \infty, \ (r \gtrsim r_{v|n|}).
\end{array}\right.
\end{eqnarray}
\cite{VS,HK}. For NO strings, Eqs. (\ref{eq:f_lo})-(\ref{eq:atheta_lo}) may be used to estimate string parameters, such as the tension, to within numerical factors of order unity. In order to calculate the constants of motion for the string described by the ansatz (\ref{eq:ansatz_phi})-(\ref{eq:ansatz_Az*}), we must solve Eq. (\ref{NO_vector+}), at least approximately within each, qualitatively different, region of the string core in order to obtain a similar expression for the function $a(r)$. Let us now assume that the external charge density described by the function $j(r)$ in Eq. (\ref{j_zj_0}) is localized in a small region close to the string core, i.e. 
\begin{eqnarray} \label{j(r)}
j(r) = \left \lbrace
\begin{array}{rl}
k,& \ 0 \leq r \leq \delta_{|n|} \\
0,& \  r \geq \delta_{|n|},
\end{array}\right.
\end{eqnarray}
where $\delta_{|n|} \leq r_{s|n|}$ is a length-scale, which may also depend on the value of $|n|$, but which is otherwise arbitrary and $k \in \mathbb{R}$ is a constant, representing the two-dimensional charge density, with dimensions $[l]^{-2}$. We now obtain approximate solutions for $a(r)$ in each of the four regions $0 \leq r \leq \delta_{|n|}$, $\delta_{|n|} \leq r \leq r_{s|n|}$, $r_{s|n|} \leq r \leq r_{v|n|}$ and $r_{v|n|} \leq r < \infty$. For the first region, $r \in [0,\delta_{|n|}]$, we use the substitution
\begin{eqnarray} \label{subs}
\alpha(r) = 1-a(r),
\end{eqnarray}
together with Eqs. (\ref{eq:f_lo}) and (\ref{j(r)}) in order to obtain the approximate EOM;
\begin{eqnarray} \label{NO_vector+A}
\frac{d^2\alpha}{dr^2} + \frac{1}{r}\frac{d\alpha}{dr} - \kappa^2 r^{2|n|}\alpha = k,
\end{eqnarray}
where we have defined
\begin{eqnarray} \label{kappa^2}
\kappa^2 = \frac{2}{(r_{s|n|})^{2|n|}r_v^2}.
\end{eqnarray}
The solutions to the the homogeneous equation are;
\begin{eqnarray} \label{homog_solnA}
\alpha_1(r) &=& I_0(br^{|n|+1}) \approx 1 + \mathcal{O}(b^2 r^{2|n|+2}),
\nonumber\\
\alpha_2(r) &=& K_0(br^{|n|+1}) \approx -\ln(br^{|n|+1}) + \mathcal{O}(1),
\end{eqnarray}
where $I_0$ and $K_0$ are modified Bessel functions of the first and second kinds, respectively, and the constant $b$ is defined as
\begin{eqnarray} \label{b}
b = \frac{\kappa}{|n|+1}.
\end{eqnarray}
The particular solution, $\alpha_p(r)$, is obtained via the formula
\begin{eqnarray} \label{Part_solnA}
\alpha_p(r) = \alpha_2(r) \int \frac{\alpha_1(r) k}{w(r)} dr -  \alpha_1(r) \int \frac{\alpha_2(r) k}{w(r)} dr,
\end{eqnarray}
where $w(r) = \alpha_1(r)\alpha_2'(r)-\alpha_2(r)\alpha_1'(r)$ is the Wronksian. Using the identity
\begin{eqnarray} \label{Identity}
I_{p}(br^{q})K_{p+1}(br^{q}) + I_{p+1}(br^{q})K_{p}(br^{q}) = \frac{r^{-q}}{b},
\end{eqnarray}
we see that the Wronksian takes an especially simple form;
\begin{eqnarray} \label{Wronk}
w(r) = -b(|n|+1)r^{|n|}[I_0(br^{|n|+1})K_1(br^{|n|+1}) + I_1(br^{|n|+1})K_0(br^{|n|+1})] = -\frac{|n|+1}{r}, 
\end{eqnarray}
so that the indefinite integrals in Eq. (\ref{Part_solnA}) become,
\begin{eqnarray} \label{Indef_Int}
\alpha_2(r) \int \frac{\alpha_1(r) k}{w(r)} dr &=& -\frac{k}{|n|+1} K_0(br^{|n|+1}) \int I_0(br^{|n|+1}) r dr 
\nonumber\\
&\approx& \frac{k}{2|n|+2} r^2 \ln(br^{|n|+1}),
\nonumber\\
\alpha_1(r) \int \frac{\alpha_2(r) k}{w(r)} dr &=& -\frac{k}{|n|+1} I_0(br^{|n|+1}) \int K_0(br^{|n|+1}) r dr 
\nonumber\\
&\approx&  \frac{k}{2|n|+2} r^2 \ln(br^{|n|+1}) - \frac{1}{4}kr^2,
\end{eqnarray}
where the expressions on the right-hand-side are approximately valid for small $r$. The particular solution to Eq. (\ref{NO_vector+A}) is therefore given by $\alpha_p(r) \approx \frac{1}{4}k r^2$,
and the general solution for the original function, $a(r)$, in the region $r \in [0,\delta_{|n|}]$, which we will label $a^{(1)}(r)$ is;
\begin{eqnarray} \label{a(1)}
a^{(1)}(r) \approx 1- \frac{1}{4}k r^2 - c_1I_0(br^{|n|+1}) - c_2K_0(br^{|n|+1}).
\end{eqnarray}
Finally, imposing the boundary condition at $r=0$ implies $c_1=1$ and $c_2=0$ so that, again using the small-$r$ expansions from Eq. (\ref{homog_solnA}), this simplifies to 
\begin{eqnarray} \label{Part_SolnA*}
a^{(1)}(r) \approx -\frac{1}{4}k r^2,
\end{eqnarray}
which is simply the leading order solution obtained by substituting an ansatz of the form $a(r) = p r^q$ into Eq. (\ref{NO_vector+A}) and is therefore analogous to the expressions for $f(r)$ and $a_{\theta}(r)$ given in Eqs. (\ref{eq:f_lo})-(\ref{eq:atheta_lo}). Hence $k<0$ is required in order for $a^{(1)}(r)>0$ within $r \in [0,\delta_{|n|}]$. In this case, as we shall see explicitly in the following calculations, $a(r)$ is monotonic, with no local maxima or minima between the values stipulated by the boundary conditions, $a(0)=0$ and $a(r\rightarrow \infty) \rightarrow 1$. In Sect.  \ref{String energy and momentum} we will also see that negative $k$ is also required to ensure that the additional charged matter within the string makes a positive contribution to the energy, which is a reasonable physical condition. For positive $k$ it is impossible to construct a solution with continuous first derivatives and such a solution (if it existed) would lead to negative energy contributions to the Hamiltonian while, for $k=0$, the function $a(r)$ diverges as $r \rightarrow 0$.  We therefore regard $k \geq 0$ as unphysical. 
\\ \indent 
In the region $r \in [\delta_{|n|}, r_{s|n|}]$ the EOM in $\alpha(r)$ is simply the homogenous version of Eq. (\ref{NO_vector+A}) already considered, so that 
\begin{eqnarray} \label{a(2)}
a^{(2)}(r) \approx 1- c_3I_0(br^{|n|+1}) - c_4K_0(br^{|n|+1}),
\end{eqnarray}
and in the region  $r \in [r_{s|n|}, \infty)$, it is approximately
\begin{eqnarray} \label{NO_vector+B}
\frac{d^2\alpha}{dr^2} + \frac{1}{r}\frac{d\alpha}{dr} - \gamma^2 \alpha = 0,
\end{eqnarray}
where 
\begin{eqnarray} \label{gamma^2}
\gamma^2 = \frac{2}{r_v^2} = (r_{s|n|})^{2|n|}\kappa^2,
\end{eqnarray}
giving
\begin{eqnarray} \label{a(3)}
a^{(3)}(r) \approx 1- c_5I_0(\gamma r) - c_6K_0(\gamma r).
\end{eqnarray}
However, the boundary condition $a(r\rightarrow \infty) \rightarrow 1$ requires us to set $c_5=0$, so that, relabelling the remaining (unfixed) constants, the full solution is;
\begin{eqnarray} \label{a_lo}
a(r) \approx \left \lbrace
\begin{array}{rl}
a^{(1)} = -\frac{1}{4}k r^2, & \ 0 \leq r \leq \delta_{|n|} \\
a^{(2)} = 1- C I_0(br^{|n|+1}) - D K_0(br^{|n|+1}), &  \  \delta_{|n|} \leq r \leq  r_{s|n|} \\ 
a^{(3)} = 1 - BK_0(\gamma r), & \  r \geq r_{s|n|}.
\end{array}\right.
\end{eqnarray}
In order to construct an approximate solution for $a(r)$, precisely analogous to those for $f(r)$ and $a_{\theta}(r)$ given in Eqs. (\ref{eq:f_lo}) -(\ref{eq:atheta_lo}), we may also choose to divide the region $r \in [r_{s|n|}, \infty)$ into two, namely $r \in [r_{s|n|}, r_{s|n|}]$ and  $r \in [r_{v|n|}, \infty)$, where we approximate $a(r)$ in the latter as $a^{(4)}(r) \approx 1$. For the time being though, we will not choose to do this, since we wish to demonstrate that an approximate solution with continuous first derivates, valid for the whole range $r \in [0,\infty)$, exists. In the solution (\ref{a_lo}) there are four remaining free constants, $\left\{k,C,D,B\right\}$ and we may impose the following set of boundary conditions in order to fix their values in terms of the parameter $\delta_{|n|}$;
\begin{eqnarray} \label{BCs}
a^{(1)}(\delta_{|n|}) &=& a^{(2)}(\delta_{|n|}), \ \ \ a'^{(1)}(\delta_{|n|}) = a'^{(2)}(\delta_{|n|}), \nonumber\\
a^{(2)}(r_{s|n|}) &=& a_p^{(3)}(r_{s|n|}), \ \ \ a'^{(2)}(r_{s|n|}) = a'^{(3)}(r_{s|n|}).
\end{eqnarray}
Imposing $a^{(1)}(\delta_{|n|}) = a^{(2)}(\delta_{|n|})$ allows us to write $C$ implicitly in terms of $\left\{k,D,\delta_{|n|}\right\}$, so that
\begin{eqnarray} \label{BC_1}
C = \left(1 + \frac{1}{4}k\delta_{|n|}^2\right) \left[I_0(b (\delta_{|n|})^{|n|+1}) + \frac{D}{C}K_0(b (\delta_{|n|})^{|n|+1}) \right]^{-1},
\end{eqnarray}
as does imposing the parallel condition $a'^{(1)}(\delta_{|n|}) = a'^{(2)}(\delta_{|n|})$;
\begin{eqnarray} \label{BC_2}
C = \frac{1}{2}\frac{k}{\kappa}(\delta_{|n|})^{1-|n|} \left[I_1(b (\delta_{|n|})^{|n|+1}) - \frac{D}{C}K_1(b (\delta_{|n|})^{|n|+1}) \right]^{-1}.
\end{eqnarray}
Imposing $a^{(2)}(r_{s|n|}) = a_p^{(3)}(r_{s|n|})$ allows us to fix $B$, explicitly, in terms of $\left\{k,C,D\right\}$,
\begin{eqnarray} \label{BC_3}
B = \frac{C}{K_0(\gamma r_{s|n|})} \left[I_0(b (r_{s|n|})^{|n|+1}) + \frac{D}{C} K_0(b (r_{s|n|})^{|n|+1})\right],
\end{eqnarray}
as does imposing $a'^{(2)}(r_{s|n|}) = a'^{(3)}(r_{s|n|})$;
\begin{eqnarray} \label{BC_4}
B = -\frac{C}{K_1(\gamma r_{s|n|})}\left[I_1(b (r_{s|n|})^{|n|+1}) - \frac{D}{C} K_1(b (r_{s|n|})^{|n|+1})\right],
\end{eqnarray}
where we have used the fact that $\kappa (r_{s|n|})^{|n|}/\gamma = 1$. Equating the two expressions for $B$, Eqs. (\ref{BC_3})-(\ref{BC_4}), then allows us to fix the ratio $D/C$ entirely in terms of known constants, 
\begin{eqnarray} \label{D/C}
\frac{D}{C} = -\left[\frac{I_0(b (r_{s|n|})^{|n|+1})}{K_0(\gamma r_{s|n|})}+\frac{I_1(b (r_{s|n|})^{|n|+1})}{K_1(\gamma r_{s|n|})}\right]\left[\frac{K_0(b (r_{s|n|})^{|n|+1})}{K_0(\gamma r_{s|n|})}-\frac{K_1(b (r_{s|n|})^{|n|+1})}{K_1(\gamma r_{s|n|})}\right]^{-1},
\end{eqnarray}
and equating the two expressions for $C$, Eqs. (\ref{BC_1})-(\ref{BC_2}), fixes $k$ in terms of $\delta_{|n|}$. The final result is;
\begin{eqnarray} \label{k}
k(\delta_{|n|}) = \left[\frac{1}{2} \kappa^{-1} (\delta_{|n|})^{1-|n|} \frac{[I_0(b(\delta_{|n|})^{|n|+1}) + (D/C) K_0(b(\delta_{|n|})^{|n|+1})]}{[I_1(b(\delta_{|n|})^{|n|+1}) - (D/C) K_1(b(\delta_{|n|})^{|n|+1})]}-\frac{1}{4}\delta_{|n|}^2\right]^{-1}. 
\end{eqnarray}
For a given value of $\delta_{|n|}$ (together with fixed values of $r_{s|n|}$ and $r_{v|n|}$, determined by the fundamental model parameters, $\eta$, $\lambda$ and $e$ together with chosen values of the phenomenological parameters $\epsilon$ and $\sigma$), we can then determine the values of each of the remaining parameters $\left\{k,C,D,B\right\}$ via back substitution, to obtain a complete solution for $a(r)$. 
\\ \indent
However, since we are interested only in obtaining order of magnitude estimates of the string constants of motion, we would like some way of estimating the values of the constants above using simpler, but approximate, expressions in terms of the more fundamental parameters of the theory. We may estimate the value of the ratio $D/C$ by writing out the arguments of the Bessel functions in Eq. (\ref{D/C}) explicitly, so that
\begin{eqnarray} \label{approxs}
b (r_{s|n|})^{|n|+1} &=& \frac{\sqrt{2}}{|n|+1}\left(\frac{r_{s|n|}}{r_v}\right) \approx \sqrt{2\beta} \frac{|n|^{\sigma}}{|n|+1},
\nonumber\\
\gamma r_{s|n|} &=& \sqrt{2} \left(\frac{r_{s|n|}}{r_v}\right)  \approx \sqrt{2\beta} |n|^{\sigma},
\end{eqnarray}
where the parameter $\beta$ is defined, as in the usual literature, via
\begin{eqnarray}
\beta =  \left(\frac{r_{s}}{r_v}\right)^2.
\end{eqnarray}
If we now assume critical coupling, $\beta \approx 1$, together with the condition
\begin{eqnarray} \label{sigma}
\sigma = \epsilon = \frac{1}{2},
\end{eqnarray}
we have
\begin{eqnarray} \label{D/C*}
\frac{D}{C} \approx -\left[\frac{I_0(\sqrt{|n|}/(|n|+1))}{K_0(\sqrt{|n|})}+\frac{I_1(\sqrt{|n|}/(|n|+1))}{K_1(\sqrt{|n|})}\right]\left[\frac{K_0(\sqrt{|n|}/(|n|+1))}{K_0(\sqrt{|n|})}-\frac{K_1(\sqrt{|n|}/(|n|+1))}{K_1(\sqrt{|n|})}\right]^{-1}.
\end{eqnarray}
Though the assumption (\ref{sigma} ) may seem arbitrary at this point, we will see in the following section that it allows a correspondence between the constants of motion for the field-theoretic strings, considered here, and those of the $F$-strings, considered in Sects. \ref{Wound-strings in the KS geometry}-\ref{String loops with uniform, geodesic windings}, to be drawn. Eq. (\ref{D/C*}) may be evaluated numerically for any value of $|n| \in [1,\infty)$ and a sample of results (evaluated to two decimal places), is given below;
\begin{eqnarray}
\left(\frac{D}{C}\right)_{|n|=1} &=& 5.31, \ \ Ê \left(\frac{D}{C}\right)_{|n|=10} = 0.83, \ \ Ê\left(\frac{D}{C}\right)_{|n|=10^2} = 0.15,
\nonumber\\
\left(\frac{D}{C}\right)_{|n|=10^3} &=& 0.04, Ê \left(\frac{D}{C}\right)_{|n|=10^4} = 0.01,  Ê\lim_{|n| \rightarrow\infty} \left(\frac{D}{C}\right) = 0^-.
\end{eqnarray}
For ``reasonable" values of $|n|$, $|n| \sim \mathcal{O}(1)-\mathcal{O}(10)$, it is reasonable to estimate the ratio $D/C$ as
\begin{eqnarray} \label{D/C**}
\frac{D}{C} \sim \mathcal{O}(1).
\end{eqnarray}
Since the length-scale $\delta_{|n|}$ is at most of the order of the overall size of the string core, we have that
\begin{eqnarray} \label{delta<<r_s}
\delta_{|n|} \lesssim r_{s|n|} \leq r_{v|n|},
\end{eqnarray}
and we may obtain a crude estimate of  $k(\delta_{|n|})$ by substituting Eq. (\ref{D/C**}) into Eq. (\ref{k}) and expanding the Bessel functions to first order. This gives,
\begin{eqnarray} \label{k*}
k(\delta_{|n|}) \approx \frac{2|n|+2}{\delta_{|n|}^2}\ln(b(\delta_{|n|})^{|n|+1})^{-1},
\end{eqnarray}
so that $k<0$, as expected. Substituting both (\ref{D/C**}) and (\ref{k*}) into either of the expressions for $C$, Eqs. (\ref{BC_1})-(\ref{BC_2}), then yields 
\begin{eqnarray} \label{C=D}
C \approx D \approx -\frac{1}{\ln(b(\delta_{|n|})^{|n|+1})}.
\end{eqnarray}
Again using Eq. (\ref{approxs}), together with Eq. (\ref{D/C**}) in Eqs. (\ref{BC_3})-(\ref{BC_4}), we then have;
\begin{eqnarray} \label{B*}
B &\approx& \frac{C}{K_0(\sqrt{n})} \left[I_0(\sqrt{n}/(|n|+1)) + K_0(\sqrt{n}/(|n|+1))\right]
\nonumber\\
&\approx& -\frac{C}{K_1(\sqrt{n})}\left[I_1(\sqrt{n}/(|n|+1)) - K_1(\sqrt{n}/(|n|+1))\right],
\end{eqnarray}
and either of these expressions imply that 
\begin{eqnarray} \label{B**}
B \approx \mathcal{O}(10) \times C,
\end{eqnarray}
for  $|n| \sim \mathcal{O}(1)-\mathcal{O}(10)$. Having obtained estimates of the values of the the constants $\left\{k,C,D,B\right\}$ in terms of the single adjustable parameter $\delta_{|n|}$, which we have assumed to be small in relation to the width of the string, (Eq. (\ref{delta<<r_s})), we are now able to calculate order of magnitude estimates for the string constants of motion, which are given in the following section.
%
\subsection{String energy and momentum} \label{String energy and momentum}
Substituting the general ansatzes for the abelian-Higgs fields, Eqs. (\ref{eq:ansatz_phi})-(\ref{eq:ansatz_Az*}), and for the components of $j^{\nu}$, Eq. (\ref{j_zj_0}), into the expressions for $T^{0}{}_{0}$,  $T^{0}{}_{z}$ and  $T^{0}{}_{\theta}$ in Eqs. (\ref{eq:H_in_cp_coords})-(\ref{eq:P^theta}) and into the expression for $J^{0}$ given by Eqs. (\ref{J})-(\ref{mathcal{J}}), yields;
\begin{eqnarray} \label{T^{0}{}_{0}*}
T^{0}{}_{0} = T^{00} &=& \eta^2 |n|^2 \left[\frac{1}{|n|^2}\left(\frac{df}{dr}\right)^2 + \frac{1}{2}\frac{r_v^2}{r^2}\left(\frac{da_{\theta}}{dr}\right)^2 + \frac{f^2}{r^2}(1-a_{\theta})^2 + \frac{1}{4|n|^2}\frac{1}{r_{s}^2}(f^2-1)^2\right]
\nonumber\\ 
&+& \eta^2 |n|^2 k_z^2 \left[r_v^2\left(\frac{da}{dr}\right)^2 + 2f^2(1-a)^2 - 2r_v^2aj\right],
\end{eqnarray}
\begin{eqnarray} \label{T^{0}{}_{z}*}
T^{0}{}_{z} = -T^{0z} = \pm \eta^2 |n|^2 k_z^2 \left[r_v^2\left(\frac{da}{dr}\right)^2 + 2f^2(1-a)^2 - 2r_v^2aj\right],
\end{eqnarray}
\begin{eqnarray} \label{T^{0}{}_{theta}*}
T^{0}{}_{\theta} = -r^2T^{0\theta} = \pm \eta^2 |n|^2 k_z \left[2f^2(1-a)(1-a_{\theta}) + r_v^2\frac{da}{dr}\frac{da_{\theta}}{dr} - r_v^2a_{\theta}j\right],
\end{eqnarray}
\begin{eqnarray} \label{T^{0}{}_{0}*}
J^{0} = J_{0} = \mp \eta^2|n| k_z \left[2f^2(1-a) + r_v^2 j\right],
\end{eqnarray}
where we have also used the dispersion relation, Eq. (\ref{disp_reln}). Using the approximate solutions for $f(r)$, $a_{\theta}(r)$ and $a(r)$, Eqs. (\ref{eq:f_bc})-(\ref{eq:alpha_bc}) and (\ref{a_lo}), respectively, together with the ansatz for $j(r)$, Eq. (\ref{j(r)}), we may integrate the expressions above over the relevant ranges  of $r$, $\theta$ and $z$, to obtain expressions for the constants of motion. Taking $C \approx D$, as in Eq. (\ref{D/C**}), rewriting $k_z = 2\pi/\lambda_z$ as in Eq. (\ref{lambda_z}) and denoting the string length via $\Delta = |z_f-z_i|$, this gives;
\begin{eqnarray} \label{P_0}
P_0 = P^0 &\approx& 2\pi \eta^2 |n| \Delta + 2\pi \eta^2 |n|^{2-2\epsilon} \Delta + 2\pi \eta^2 |n|^2 \ln\left(\sqrt{\beta_{|n|}}\right) \Delta + \frac{\pi}{4}\eta^2 |n|^{2\sigma} \Delta
\nonumber\\
&+& 2\pi \eta^2 |n| \times \frac{(2\pi)^2r_{v|n|}^2}{\lambda_z^2} \Delta \left[\frac{1 + |n|(\beta_{|n|}-1)}{\beta_{|n|}}\right]
\nonumber\\
&+& 2\pi \eta^2 |n|^{2-2\epsilon} \times \frac{(2\pi)^2r_{v|n|}^2}{\lambda_z^2} \Delta \left[\frac{1}{16}k^2\delta_{|n|}^2 + D^2|n|^2\ln\left(\frac{r_{s|n|}}{\delta_{|n|}}\right) + B \ln\left(\sqrt{\beta_{|n|}}\right)\right] 
\nonumber\\
&+& \frac{\pi}{4}\eta^2|n|^{2-2\epsilon} \times  \frac{(2\pi)^2r_{v|n|}^2}{\lambda_z^2}k^2\delta_{|n|}^4,
\end{eqnarray}
\begin{eqnarray} \label{P_z}
P_z = -P^z \approx \pm 2\pi \eta^2 |n| \times \frac{(2\pi)^2r_{v|n|}^2}{\lambda_z^2} \Delta \left[\frac{1 + |n|(\beta_{|n|}-1)}{\beta_{|n|}}\right],
\end{eqnarray}
\begin{eqnarray} \label{l_{theta}}
l_{\theta} &\approx& \pm 2\pi \eta^2 |n| \times \frac{(2\pi)^2r_{v|n|}^2}{\lambda_z} \Delta \left[\frac{1 + |n|(\beta_{|n|}-1)}{\beta_{|n|}}\right],
\nonumber\\
l^{\theta} &\approx& \mp (2\pi)^2\eta^2|n|\frac{\Delta}{\lambda_z}\left[1 + 2|n|\ln\left(\sqrt{\beta_{|n|}}\right)\right] \pm (2\pi)^2 \eta^2|n|^{2-2\epsilon} \times \frac{1}{2}k\delta_{|n|}^2\frac{\Delta}{\lambda_z}
\nonumber\\ 
&\mp& (2\pi)^2\eta^2|n|^{2-2\epsilon}\frac{\Delta}{\lambda_z}\left[-\frac{1}{2}k\delta_{|n|}^2 + 2|n|D\ln\left(\frac{r_{s|n|}}{\delta_{|n|}}\right) + B\ln\left(\sqrt{\beta_{|n|}}\right)\right],
\end{eqnarray}
\begin{eqnarray}  \label{Q}
Q \approx (2\pi)^2\eta^2 |n| r_{v|n|}^2  \frac{\Delta}{\lambda_z} e\left[1 + \frac{1}{2}|n|^{-2\epsilon}\delta_{|n|}^2k - \frac{|n|}{|n|+1}\beta_{|n|}\right],
\end{eqnarray}
where we have also used the additional approximation
\begin{eqnarray} \label{n_approx}
|n| \approx |n|+1, \ \ \ |n| \geq 1, 
\end{eqnarray}
except in the expression for $Q$, where the distinction is important for $|n| \sim \mathcal{O}(1)$, and where the parameter $\beta_{|n|}$ is defined according to
\begin{eqnarray} \label{beta_n}
\beta_{|n|} = \left(\frac{r_{v|n|}}{r_{s|n|}}\right)^2 = |n|^{2(\epsilon-\sigma)}\beta. 
\end{eqnarray}
Next, we notice that the terms involving $D$, $-k\delta_{|n|}^2$ and $-k\delta_{|n|}^4$ in the expressions above may be neglected if
\begin{eqnarray} 
C \approx D \approx -\frac{1}{\ln((b\delta_{|n|})^{|n|+1})} \lesssim \frac{1}{|n|\ln(r_{s|n|}/\delta_{|n|})},
\end{eqnarray}
which requires
\begin{eqnarray} \label{delta_bound}
\delta_{|n|} \gtrsim \frac{1}{\sqrt{2}}\sqrt{\beta_{|n|}}|n|^{-\epsilon}r_{s|n|}.
\end{eqnarray}
Setting $\sigma = \epsilon = 1/2$ as in Eq. (\ref{sigma}) and assuming critical coupling ($\beta=1$), so that 
\begin{eqnarray} 
r_{v|n|} = r_{s|n|} = r_{c|n|}, \ \ \ (\beta_{|n|}=1),
\end{eqnarray}
the bound (\ref{delta_bound}) is marginally consistent with Eq. (\ref{delta<<r_s}) for $|n| \sim \mathcal{O}(1)$, but easily consistent for $|n| > \mathcal{O}(1)$. Under these conditions, the formulae for the constants of motion simplify dramatically, giving;
\begin{eqnarray} \label{P_0*}
P_0 = P^0 \approx 2\pi \eta^2 |n| \Delta \left(1 + \frac{(2\pi)^2r_{c|n|}^2}{\lambda_z^2}\right),
\end{eqnarray}
\begin{eqnarray} \label{P_z*}
P_z = -P^z \approx \pm 2\pi \eta^2 |n| \times \frac{(2\pi)^2r_{c|n|}^2}{\lambda_z^2} \Delta,
\end{eqnarray}
\begin{eqnarray} \label{l_{theta}*}
l_{\theta} \approx -r_{c|n|}^2l^{\theta}  \approx \pm (2\pi)^2\eta^2 |n| \frac{r_{c|n|}^2}{\lambda_z} \Delta.
\end{eqnarray}
\begin{eqnarray}  \label{Q*}
Q \approx (2\pi)^2\eta^2 \frac{ r_{c|n|}^2}{\lambda_z} \Delta e,
\end{eqnarray}
It is interesting that under these, quite natural, conditions the contribution to the string momenta, $P_z$ and $l_{\theta}$, and to the total charge, $Q$, from the externally coupled charged fluid is at most comparable and very likely sub-dominant to the contribution from the vacuum part of the solution; that is, from the configuration of the abelian-Higgs fields. In this case we also have that 
\begin{eqnarray}  \label{Q=(e/n)l_{theta}}
Q \approx \frac{e}{|n|}l_{\theta}.
\end{eqnarray}
However, in principle we may fine-tune the values of the parameters such that $-\frac{1}{2}|n|^{-2\epsilon}\delta_{|n|}k \sim \mathcal{O}(1)$,  $Q \approx 0$, to obtain charge neutral strings and the order of magnitude values for the remaining constants of motion remain unaffected.
\\ \indent
Next, we consider the case where the section of string we are investigating, $\Delta=z_f-z_i$, is equal to an integer multiple, $m$, of the length-scale, $\lambda_z$, which is the characteristic length-scale over which the phase along a line of constant $(r,\theta)$ varies by $\pm 2\pi |n|$;
\begin{eqnarray}
\Delta = m\lambda_z, \ \ \ m \in \mathbb{Z},
\end{eqnarray}
so that the expressions for the conserved momenta and the electric charge may then be written in the form,
\begin{eqnarray} \label{P_z**}
P_{z} = -P^{z} \approx  \pm 2\pi \eta^2 |n|m \times \frac{(2\pi)^2r_{c|n|}^2}{\lambda_z},
\end{eqnarray}
\begin{eqnarray}  \label{l_{theta}**}
l_{\theta} \approx -r_{c|n|}^{2}l^{\theta}  \approx \pm (2\pi)^2\eta^2 |n| m r_{c|n|}^2,
\end{eqnarray}
\begin{eqnarray}  \label{Q**}
Q \approx (2\pi)^2\eta^2 |n| m r_{c|n|}^2 e
\end{eqnarray}
According to the construction above, we may interpret $\lambda_z$ as the distance (along the $z$-direction), over which the ``overall" orientation of the vortex-slices which compose the string rotate by $\pm2\pi$. The positive integer $|m|$ then gives the number of such windings and the sign of $m$ determines their sense. Though individual vortices do not have an overall orientation, and phase-shifting by a constant value is simply a gauge transformation, in the ansatz we have constructed the phase along a line of fixed $(r,\theta)$ changes by $\pm 2\pi |n|$ over $\lambda_z$ at any time $t$. In order to express this, we may therefore imagine that each $|n|$-vortex slice possesses an ``overall" orientation which rotates by $\pm 2\pi$ over the same distance. Physically speaking, $|m|$ is also equal to the number of windings (along the total length of the string, $\Delta$), of the magnetic flux lines around the central axis of the vortex core. Their sense is the same as the windings in the phase.
\\ \indent
Finally, the total conserved energy, linear momentum and ``angular momentum", $E_{|n|}$, $P_{|n|}$ and $l_{|n|}$, and the total conserved $4$-momentum, $\Pi_{|n|}$, are:
\begin{eqnarray} \label{E_n}
E_{|n|} \approx 2\pi \eta^2 |n|\Delta \left(1+\frac{(2\pi)^2r_{c|n|}^2}{\lambda_z^2}\right),
\end{eqnarray}
\begin{eqnarray} \label{P_n}
P_{|n|} \approx 2\pi \eta^2 |n| m \times \frac{(2\pi)^2r_{c|n|}^2}{\lambda_z},
\end{eqnarray}
\begin{eqnarray} \label{l_n}
l_{|n|} \approx  (2\pi)^2 \eta^2 |n| m r_{c|n|},
\end{eqnarray}
\begin{eqnarray} \label{Pi_n}
\Pi_{|n|} \approx 2\pi \eta^2 |n| \Delta \sqrt{1 + \frac{(2\pi)^2 r_{c|n|}^2}{\lambda_z^2}}.
\end{eqnarray}
\section{Transition to loops in the field theory} \label{Transition to loops in the field theory}
We now attempt to model the chopping-off of a topological defect string with twisted flux lines from a long-string network, and the subsequent loop dynamics, using the techniques outlined at the end of Sect. \ref{String loops with uniform, geodesic windings} . Essentially, we assume that the macroscopic loop dynamics are independent of the detailed structure of the vortex core, and are determined only by the macroscopic conserved quantities (i.e. the momentum and energy) of the string.  
\\ \indent
Specifically, we do not attempt to find an ansatz, satisfying the fundamental abelian-Higgs EOM, which simultaneously describes the dynamics we obtain by appealing to the conservation of energy and momentum. Such a project is clearly worthwhile, but beyond the scope of this paper, and we instead make use of the fact that, whatever their precise structure, the scalar and gauge fields evolve so as to keep $E_{|n|}$, $l_{|n|}$ and $m$ constant. Though we initially assume $m$ to be stabilized by the conservation of $E_{|n|}$ and $l_{|n|}$, explicit stability conditions are proposed in Sect. \ref{Stability analysis}, based on considering equivalent conditions for the stability of $n_{\phi}$ in terms of $E_{s}$ and $l_{s}$. We begin by identifying
\begin{eqnarray} \label{ident*}
\Delta &\rightarrow& \frac{2\pi \rho(t_i)}{\sqrt{1-\dot{\rho}^2(t_i)}} \rightarrow \frac{2\pi \rho(t)}{\sqrt{1-\dot{\rho}^2(t)}},
\nonumber\\
\lambda_z &\rightarrow& \lambda_z(t_i) = \frac{2\pi \rho(t_i)}{\sqrt{1-\dot{\rho}^2(t_i)}m} \rightarrow {\sqrt{1-\dot{\rho}^2(t)}\lambda_z(t) =  \frac{2\pi \rho(t)}m},
\end{eqnarray}
which is the analogue of Eq. (\ref{ident}), and
\begin{eqnarray} \label{identify*}
l_{\sigma} = 2\pi \rho(t_i)P_{z}, \ \ \ l^{\sigma}(t_i) = \frac{P^{z}}{2\pi \rho(t_i)},
\end{eqnarray}
which is the analogue of (\ref{identify}), again imposing the initial condition
\begin{eqnarray} \label{bc**}
\dot{\rho}(t_i) = 0.
\end{eqnarray}
We then assume that the initial values of $E_{|n|}=P_0$, $P_z$ and $l_{\theta}$ are given by (\ref{P_0*}), (\ref{P_z**}) and (\ref{l_{theta}**}). This yields
\begin{eqnarray} \label{E_|n|}
E_{|n|} \approx (2\pi)^2 \eta^2 |n| \rho(t_i) \left(1+\frac{m^2 r_{c|n|}^2}{\rho^2(t_i)}\right) = (2\pi)^2 \eta^2 |n| \frac{\rho(t)}{\sqrt{1-\dot{\rho}^2(t)}} \left(1+\frac{m^2 r_{c|n|}^2}{\rho^2(t)}\right),
\end{eqnarray}
\begin{eqnarray} \label{l_sigma}
l_{\sigma} = -\rho^2(t_i)l^{\sigma} \approx \pm (2\pi)^3 \eta^2 |n| m^2 r_{c|n|}^2.
\end{eqnarray}
From (\ref{E_|n|}) we can determine the EOM of the string;
\begin{eqnarray} \label{EOM1}
\dot{\rho}^2 - 1 + \left(\frac{\rho^2(t_i)}{\rho^2(t_i) + m^2r_{c|n|}^2}\right)^2\left(\frac{\rho}{\rho(t_i)} + \frac{m^2r_{c|n|}^2}{\rho(t_i)\rho}\right)^2 = 0,
\end{eqnarray}
and defining the parameters $\omega_{\theta}^2(t)$ and  $\omega_{\sigma}^2(t)=1-\omega_{\theta}^2(t)$, via
\begin{eqnarray} \label{omega_{sigma}(t)*}
\omega_{\sigma}^2(t) = \frac{\rho^2(t_i)}{\rho^2(t_i)+m^2r_{c|n|}^2},
\end{eqnarray}
\begin{eqnarray} \label{omega_{theta}(t)}
\omega_{\theta}^2(t) &=& \frac{m^2r_{c|n|}^2}{\rho^2(t_i)+m^2r_{c|n|}^2},
\end{eqnarray}
by analogy with (\ref{omega_{phi}(t)}) and  (\ref{omega_{sigma}(t)}), allows us to write the solution of (\ref{EOM1}), in a convenient form,
\begin{eqnarray} \label{loop_EOM1*}
\rho(t) = \rho(t_i)\sqrt{1 + \left(\frac{1-2\omega_{\sigma}^2(t_i)}{\omega_{\sigma}^4(t_i)}\right)\sin^2\left(\frac{\omega_{\sigma}^2(t_i)}{\rho(t_i)}(t-t_i)\right)},
\end{eqnarray}
which is clearly equivalent to (\ref{loop_EOM_soln1}) under the identifications $\rho(t) \leftrightarrow \rho_s(t)$, $\omega_{\sigma}^2(t) \leftrightarrow \omega_{s\sigma}^2(t)$. 
By analogy with (\ref{crit_values}), the critical values of $\rho(t)$, $\lambda_z(t)$ and $\omega_{\sigma}^2(t)$ are 
\begin{eqnarray} \label{crit_values*}
\rho_{(c1)} &=& \rho(t_i), \ \ \ \rho_{(c2)}=\frac{1-\omega_{\sigma}^2(t_i)}{\omega_{\sigma}^2(t_i)} \rho(t_i),
\nonumber\\
\lambda_{z(c1)} &=& 2\pi\frac{\omega_{\sigma}(t_i)}{\sqrt{1-\omega_{\sigma}^2(t_i)}}r_{c|n|}, \ \ \ \lambda_{z(c2)} = 2\pi\frac{\sqrt{1-\omega_{\sigma}^2(t_i)}}{\omega_{\sigma}(t_i)}r_{c|n|},
\nonumber\\
\omega_{\sigma(c1)}^2 &=& \omega_{\sigma}^2(t_i), \ \ \ \omega_{\sigma(c2)}^2 = 1- \omega_{\sigma}^2(t_i),
\end{eqnarray}
which also implies $\lambda_z(t) \leftrightarrow d_{\phi}(t)$ and $r_{c|n|} \leftrightarrow R$. Finally, taking $\lambda_z(t) \leftrightarrow d_{\phi}(t)$ and comparing (\ref{d_{phi}2}) with the definition of $\lambda_z(t)$ in (\ref{ident*}), we have that $m \leftrightarrow n_{\phi}/a$. The correspondences between field theory and string theory parameters are discussed at greater length in the following Sect. For now, we note that the total angular momentum, $l_{|n|}$, and $4$-momentum, $\Pi_{|n|}$,  for the field-theoretic string loop are   
\begin{eqnarray} \label{l_n*}
l_{|n|} = (2\pi)^2 \eta^2 |n| m r_{c|n|}\sqrt{1 + \frac{( 2\pi)^2 r_{c|n|}^2}{\lambda_z^2(t_i)}},
\end{eqnarray}
\begin{eqnarray} \label{Pi_n*}
\Pi_{|n|} = (2 \pi)^2 \eta^2 |n|  \rho(t_i) \sqrt{1 + \frac{( 2\pi)^2r_{c|n|}^2}{\lambda_z^2(t_i)}}.
\end{eqnarray}
\section{Correspondence between wound-string and field-theoretic model parameters} \label{Correspondence between wound-string and field-theoretic model parameters}
Comparing our results for $F$-strings with uniform, geodesic windings, presented in Sects  \ref{Long, straight strings with uniform geodesic windings}-\ref{String loops with uniform, geodesic windings}, with those obtained for the modification of the NO ansatz in the abelian-Higgs model, presented in Sect.  \ref{Field configuration for a long, straight string with twisted flux-lines: Model 1}, we see that 
\begin{eqnarray} \label{correspondences1}
E_s/a \leftrightarrow E_{|n|}, \ \ \ P_s/a \leftrightarrow P_{|n|}, \ \ \  l_s/a \leftrightarrow l_{|n|}, \ \ \ \Pi_s/a \leftrightarrow \Pi_{|n|}
\end{eqnarray}
under the identifications
\begin{eqnarray} \label{ident_Teta}
 \mathcal{T} \leftrightarrow 2\pi \eta^2|n|.
\end{eqnarray}
\begin{eqnarray} \label{ident_Delta}
\Delta_s \leftrightarrow \Delta, \ \ \ \rho_s(t) \leftrightarrow \rho(t),
\end{eqnarray}
\begin{eqnarray} \label{ident_dlambda}
d_{\phi} \leftrightarrow \lambda_z, \ \ \  d_{\phi}(t) \leftrightarrow \lambda_z(t),
\end{eqnarray}
\begin{eqnarray} \label{ident_rR}
R \leftrightarrow r_{c|n|},
\end{eqnarray}
\begin{eqnarray} \label{ident_nm}
n_{\phi}/a \leftrightarrow m.
\end{eqnarray}
However, since the wound-string models live in warped space, whereas the field-theoretic strings live in ordinary Minkowski space, we are automatically comparing strings of different lengths, namely of length $a\Delta_s$ or $a\rho_s(t)$ in the former and $\Delta$ or $\rho(t)$ in the latter. Comparing loops of equal size, all the relevant quantities become exactly equivalent. Since $T \sim \alpha'^{-1}$, we may also rewrite (\ref{ident_Teta}) as
\begin{eqnarray} \label{ident_Teta*}
a^{\xi}\sqrt{\alpha'}^{-1} \leftrightarrow \eta, \ \ \ a^{-2\xi} \leftrightarrow |n|.
\end{eqnarray}
For $\xi = 0$, the symmetry-breaking energy scale is equal to the fundamental string energy-scale and defect strings with $|n|\sim\mathcal{O}(1)$ are energetically and dynamically equivalent to wound $F$-strings, whereas for $\xi = 1$, the symmetry-breaking energy scale is equivalent to the the warped-string energy-scale and defects with $|n| \sim \mathcal{O}(a^{-2})$ are required. In the next subsection, we consider this point in relation to dynamical models of string-winding and phase/flux-winding formation in each scenario, respectively.
\subsection{Specific, but natural, models for winding formation} \label{Specific, but natural, models for winding formation}
For an $F$-string, in warped space, with windings which form dynamically due to the motion of the string in the compact space, we expect the number of windings in a loop of initial size $a\rho_s(t_i)$ to be given by 
\begin{eqnarray} \label{n_dyn1}
n_{\phi} = \frac{v_{4D}(t_i)}{v_{E}(t_i)}\frac{a\rho_s(t_i)}{R},
\end{eqnarray}
where $v_{4D}(t_i)$ is the velocity of the ``end point" of the string, that is, of the ``point" on the string, as seen by an observer comoving with the string CoM, lying instantaneously at the horizon in the large dimensions, and $v_{E}(t_i)$ is its velocity in the compact directions \cite{Cycloops,Thesis}. Before the string chops-off from the network, the two velocities are related via
\begin{eqnarray} \label{velocities1}
v_{4D}^2(t \leq t_i) + v_{E}^2(t \leq t_i) = 1, 
\end{eqnarray}
where we have also assumed that $v_{4D}^2(t \leq t_i)=const$,  $v_{E}^2(t \leq t_i)=const$ in order to obtain the relation (\ref{n_dyn1}). Since Neumann boundary conditions in string theory mean that the end point of the string at the horizon must move with resultant velocity $c=1$, in a warped space model this naturally implies
\begin{eqnarray} \label{velocities2}
v_{4D}(t_i)= a, \ \ \ v_{E}(t_i) = \sqrt{1-a^2}.
\end{eqnarray}
Substituting this into (\ref{n_dyn1}) gives
\begin{eqnarray} \label{n_dyn2}
n_{\phi} = \frac{\sqrt{1-a^2}\rho_s(t_i)}{R},
\end{eqnarray}
and substituting this into (\ref{omega_{phi}(t)})-(\ref{omega_{sigma}(t)}) yields
\begin{eqnarray}
\omega_{s\sigma}(t_i)=a, \ \ \ \omega_{\phi}(t_i) = \sqrt{1-a^2}.
\end{eqnarray}
This also agrees with Eq. (\ref{n_phi_approx1}) and with the expressions for $d_{\phi(c1)}$ and $d_{\phi(c2)}$ in (\ref{crit_values}).
\\ \indent
If we now assume that the phase at each point on the circumference of the vortex core of a long, straight, topological defect string (at critical coupling), ``moves" with resultant velocity $c=1$, we may paramaterise its forward motion along the length of the string, $v_z(t \leq t_i)$, and its motion around the core central axis, $v_{\theta}(t \leq t_i)$, in terms of a single parameter, $\delta$, via 
\begin{eqnarray}
v_z(t \leq t_i) = \delta, \ \ \ v_{\theta}(t \leq t_i) = \sqrt{1-\delta^2}, \ \ \ 0 < \delta^2 < 1.
\end{eqnarray}
For a string of initial length $2\pi \rho(t_i)$ which chops-off from the network, the number of windings in the phase (by $2\pi$), and/or the number of windings in the magnetic flux lines is $m$, where 
\begin{eqnarray}
m = \frac{\sqrt{1-\delta^2}}{\delta} \frac{\rho(t_i)}{r_{c|n|}}.
\end{eqnarray}
However, since such a disturbance may only propagate along the length of the string with velocity $v_z(t \leq t_i) = \delta$, strictly speaking, we should consider only loops of initial size $\delta \rho(t_i)$, where $\rho(t_i) \leq t_i$. In this case, we have that  
\begin{eqnarray}
m \rightarrow m' = \delta m = \frac{\sqrt{1-\delta^2}\rho(t_i)}{r_{c|n|}}.
\end{eqnarray}
Under these circumstances, the identifications (\ref{ident_Teta})-(\ref{ident_rR}), plus
\begin{eqnarray}
a \leftrightarrow \delta,
\end{eqnarray}
yield 
\begin{eqnarray}
n_{\phi} \leftrightarrow m',
\end{eqnarray}
\begin{eqnarray}
\omega_{s\sigma}(t_i) = a \leftrightarrow \omega_{\sigma}(t_i) = \delta, \ \ \ \omega_{\phi}(t_i) = \sqrt{1-a^2} \leftrightarrow \omega_{\theta}(t_i) = \sqrt{1-\delta^2},
\end{eqnarray}
together with
\begin{eqnarray} \label{correspondences3}
E_s \leftrightarrow E'_{|n|}, \ \ \ P_s \leftrightarrow P'_{|n|}, \ \ \  l_s \leftrightarrow l'_{|n|}, \ \ \ \Pi_s \leftrightarrow \Pi'_{|n|}.
\end{eqnarray}
We may also make the identification \footnote{A similar relation, i.e. $|n| \sim a^{-\xi} \sim \delta^{-\xi} \sim \omega_{\sigma}^{-\xi}(t_i) =  \left(1-\omega_{\theta}^{2}(t_i)\right)^{-\xi/2} = \left(1+(2\pi)^2r_{c|n|}^2/\lambda_z^2(t_i)\right)^{-\xi/2} $, with $\xi=1$, was proposed in \cite{Pinch}. This correspondence was based on directly comparing $P_{0}$ and ${}^{(s)}P_{0}$, rather than $E_{|n|}=\sqrt{P_{0}P^{0}}$ and $E_s=\sqrt{{}^{(s)}P_{0}{}^{(s)}P^{0}}$, as defined in Sects. \ref{Wound-strings in the KS geometry} and \ref{Review of the abelian-Higgs Model}. Since comparing $P_{0}$ with ${}^{(s)}P_{0}$, and $P^{0}$ with ${}^{(s)}P^{0}$ give different results, it is unclear how to make valid comparisons between quantities in warped and unwarped space. However, we believe that the method outlined here now correctly incorporates the appropriate powers of the warp factor coming from the metric, leading to a modification of the earlier result.}
\begin{eqnarray} \label{ident_Teta**}
|n| \sim \frac{1}{a^{2\xi}} \sim \frac{1}{\delta^{2\xi}} \sim \frac{1}{\omega_{\sigma}^{2\xi}(t_i)} =  \frac{1}{(1-\omega_{\theta}^{2}(t_i))^{\xi}} = \left(1+\frac{(2\pi)^2r_{c|n|}^2}{\lambda_z^2(t_i)}\right)^{\xi} .
\end{eqnarray}
Dynamically, this makes sense if we consider that, in the limit $\delta \rightarrow 1$, the ``velocity" of the phase factor of field $\phi(x)$ along the length of the string is $c$, so that no ``spare" velocity exists to form windings. When $\delta \ll 1$, the field $\phi(x)$ takes a highly circuitous route from the top of the potential hill, in the phase space of the theory, to the circle of degenerate minima, over the period of string formation, $\tau_c$. Thus, in this model, the parameters $\delta$ and $\xi$ are related in some way to the dynamics of the symmetry-breaking process itself. 
\section{Stability analysis} \label{Stability analysis}
For the wound-string model presented in Sect.  \ref{Long, straight strings with uniform geodesic windings}-\ref{String loops with uniform, geodesic windings}, Blanco-Pillado and Iglesias \cite{BlancoPillado} showed that string-loop configurations with initial radius $a\rho_s(t_i)$ and number of windings $n_{\phi}$ are stabilised if the ``angular momentum" in the $\sigma$-direction, ${}^{(s)}l_{\sigma}$, exceeds a certain critical value;
\begin{eqnarray} \label{Fstab1}
{}^{(s)}l_{\sigma} > (2\pi)^2 \mathcal{T} a^2n_{\phi}^2R^2.
\end{eqnarray}
This was derived by turning on a small perturbation in the $\chi$-coordinate, which yields the following perturbation of the wound-string Lagrangian;
\begin{eqnarray} \label{ActionPert}
\delta L = \left(\frac{{}^{(s)}l_{\sigma}}{(2\pi)^2 a^2\mathcal{T}} + n_{\phi}^2R^2\right)\delta\dot{\chi}^2 - \frac{a^2}{R^2}\left(\frac{{}^{(s)}l_{\sigma}}{(2\pi)^2 a^2\mathcal{T}} - n_{\phi}^2R^2\right)\delta\chi^2.
\end{eqnarray}
For long-string sections, the equivalent result is 
\begin{eqnarray} \label{Fstab2}
{}^{(s)}P_{z} > (2\pi)^2 \mathcal{T} a^3n_{\phi}\frac{R^2}{d_{\phi}}.
\end{eqnarray}
For the boundary conditions (\ref{n_{phi}1})/(\ref{n_{phi}1A}), we have that ${}^{(s)}P_{z} = (2\pi)^2 \mathcal{T} a n_{\phi}R^2/d_{\phi}$ according to Eq. (\ref{P_z_P^z1}) and ${}^{(s)}l_{\sigma} = (2\pi)^2 \mathcal{T} n_{\phi}^2R^2$ by Eq. (\ref{l_{sigma}_l^{sigma}1}). The conditions (\ref{Fstab1}) and (\ref{Fstab2}) therefore both reduce to
\begin{eqnarray} \label{KSdefn}
a^2 < 1,
\end{eqnarray}
which is automatically satisfied for the KS geometry. According to the correspondences proposed in Sect.  \ref{Specific, but natural, models for winding formation} above, the equivalent stability conditions for the field-theory model presented in Sect.  \ref{Field configuration for a long, straight string with twisted flux-lines: Model 1} would therefore be
\begin{eqnarray} \label{Defstab1}
l_{\sigma} > (2\pi)^3 \eta^2 |n| \delta^2 m'^2 r_{c|n|}^2,
\end{eqnarray}
\begin{eqnarray}  \label{Defstab2}
P_{z} > (2\pi)^3 \eta^2 |n| \delta^3 m' \frac{r_{c|n|}^2}{\lambda_z},
\end{eqnarray}
which reduce to 
\begin{eqnarray} \label{FTdefn}
\delta^2 < 1.
\end{eqnarray}
This is satisfied by definition. In theory, we can construct an effective Lagrangian for the defect string by plugging the ansatz (\ref{eq:ansatz_phi})-(\ref{eq:ansatz_Az*}) directly into the abelian-Higgs Lagrangian (\ref{eq:action}). We then expect that perturbing the action with respect to $r_{c|n|} = r_{s|n|}=r_{v|n|}$ should yield the conditions (\ref{Defstab1})-(\ref{Defstab2}), since this is the analogue of perturbing $\chi$ in the wound-string model, which alters the effective radius of the windings according to the Hopf fibration (\ref{KSmetric}). Direct demonstration of this lies beyond the scope of this paper, though it remains important to establish whether a correspondence exists between the stability conditions in the string theory and field theory models, according to the parameter correspondences already established. 
\section{Conclusions} \label{Conclusions}
We have shown that a field configuration describing an abelian-Higgs string which does not posses cylindrical symmetry exists, so long as a small amount of externally coupled charge is present within the string core. In our model, the lines of magnetic flux, and of constant phase in the scalar field, adopt helical configurations and lead to the existence of a second winding number, $|m| \geq 0$, which quantifies the number of twists around the string central axis, in addition to the topological winding number, $|n| \geq 1$. The violation of cylindrical symmetry implies that it is physically meaningful to translate the string in the direction of its own length and this gives rise to the existence of conserved quantities, namely linear momentum in the case of long, straight strings and angular momentum in the case of string loops. For loops, the angular momentum interacts with the string tension, modifying the macroscopic loop dynamics, and three distinct dynamical regimes emerge, depending on the initial fraction of the total length of the flux-lines lying parallel to the direction of the string.
\\ \indent
For a simple choice of ansatz, the values of the energy and conserved momenta for both long strings and loops, match those of wound $F$-strings with helical configurations in the Klebanov-Strassler geometry which, for this purpose, can be taken as a generic example of an extra-dimensional background with a simply connected internal manifold. The macroscopic loop dynamics also coincide with the wound-string case, since these are determined by the constants of motion, and a set of formal correspondences between individual string theory and field theory parameters can be established. 
\\ \indent
Crucially, the solution considered here depends on the introduction of a new length-scale, $\delta_{|n|} \lesssim r_{s|n|} \leq r_{v|n|}$, which is equal to the distance (counting radially outwards from $r=0$), over which the externally coupled charged fluid is distributed within the string core, and which also determines its charge density. A small amount of such fluid is necessary to prevent divergences in the gauge field potentials as $r \rightarrow 0$. However, under reasonable assumptions, the contribution to the conserved momenta from this fluid is at most comparable and very likely sub-dominant to that arising from the configuration of the abelian-Higgs fields. In this case, explicit dependence on $\delta_{|n|}$ drops out of the final expressions for the energy and momenta. This may also be true for the contribution to the total electric charge of the string though, since the charge arising from the vacuum part of the solution necessarily has opposite sign, it is possible to tune the parameters governing the externally coupled fluid to achieve charge neutrality.
\\ \indent
The issue of stability for the resulting field configuration is key. Since the twists in the field-lines and lines of constant phase are not topologically stabilized, they must be stabilized, if at all, dynamically. The new string has higher energy than a normal, Nielsen-Olesen type defect string so that, without  the presence of a conserved momentum, energetic considerations alone would imply that the field lines should untwist as the string relaxes to a Nielsen-Olesen configuration over some characteristic timescale after formation. Though a full stability analysis has been carried out previously for the analogue case of wound-strings \cite{BlancoPillado,Cycloops,Thesis}, and the phenomenological correspondences between string theory and field theory parameters drawn here suggest equivalent relations for the twisted flux-line string, it remains an important task to verify these directly. For the sake of brevity, we have neglected to carry out such an analysis in this paper. 
\\ \indent
Finally, it is important to note that, in truth, $\delta_{|n|}$ does appear in the expressions for both the energy and conserved momenta, as well as the electric charge, and a full stability analysis should account for all possible values of the free model parameters. Most likely, the precise value of $\delta_{|n|}$ at which the field configuration is stabilized, for a given set of initial conditions, must be fixed dynamically by the stability conditions themselves, though it is not immediately clear whether a single, unique value is preferred, or whether a range of possible values exists.   
\section*{Appendix: Relation of the twisted flux line solution to traveling wave solutions for Nielsen-Olesen strings} \label{Appendix A}
The general EOM for a classical Nambu-Goto string, either open or closed, which also reproduces the correct dynamics for a Nielsen-Olesen type string in the zero-thickness approximation \cite{Anderson_book}, is well known \cite{GSW_vol1,Polchinski_vol1,Zwiebach_book}. Using the three local symmetries of the string WS (two reparameterizations and one Weyl scaling), the Nambu-Goto action may be rewritten in the form
\begin{eqnarray} \label{Nambu2}
S = -\mathcal{T}\int d^2\zeta \eta_{ab}\partial^{a}X_{I}\partial^{b}X^{I},
\end{eqnarray}
where $\eta_{ab}$ is the Minkowski metric  \cite{GSW_vol1}. The Euler-Lagrange equation derived from (\ref{Nambu2}) is simply the free, two-dimensional wave equation in the WS coordinates, $\tau$ and $\sigma$, for each embedding coordinate $X^{I}(\sigma,\tau)$,
\begin{eqnarray} \label{wave_eqn}
\left(\frac{\partial^2}{\partial \sigma^2} - \frac{\partial^2}{\partial \tau^2}\right)X^{I}(\sigma,\tau)=0.
\end{eqnarray}
For closed strings, this is sufficient to ensure that (\ref{Nambu2}) is invariant under a general variation of the embedding
\begin{eqnarray} \label{general_var}
X^{I}(\sigma,\tau) \rightarrow X'^{I}(\sigma,\tau) = X^{I}(\sigma,\tau) + \delta X^{I}(\sigma,\tau),
\end{eqnarray}
but for open strings, variation of (\ref{Nambu2}) with respect to (\ref{general_var}) gives an additional surface term
\begin{eqnarray} \label{surface_term}
-\mathcal{T}\int d\tau [X'_{I}\delta X^{I} |_{\sigma=\pi} - X'_{I}\delta X^{I} |_{\sigma=0}] = 0, 
\end{eqnarray}
which vanishes according to the open string (Dirichlet) boundary conditions. The general solution of (\ref{wave_eqn}) is the sum of two arbitrary functions in the transformed ``light-cone" WS coordinates $\sigma^{\pm} = \tau \pm \sigma$ 
\begin{eqnarray}
X^{I}(\sigma^{-},\sigma^{+}) = X^{I}_{R}(\sigma^{-}) + X^{I}_{L}(\sigma^{+}),
\end{eqnarray}
which correspond to left-moving and right-moving modes of the string. In \cite{Vachaspati&Vachaspati} the formalism for treating left and right-moving modes on Nielsen-Olesen strings (i.e. cylindrically symmetric topological defect strings of finite width), was derived from purely field-theoretic arguments, using the fundamental abelian-Higgs EOM and assuming negligible string gravity. The authors denoted the field configurations for a static local string along the $z$-axis by $\Phi(x,y)$ and $\mathcal{A}^{\mu}(x,y)$, where $(x,y,z)$ are the physical Cartesian coordinates in Minkowski space. They then substituted an ansatz of the form 
\begin{eqnarray} \label{Vachaspati_ansatz}
\phi &=& \Phi(X,Y), \nonumber\\
A^{t} &=& \dot{\psi}\mathcal{A}^{x}(X,Y) + \dot{\chi}\mathcal{A}^{y}(X,Y), \nonumber\\
A^{x} &=& \mathcal{A}^{x}(X,Y), \ \ \ A^{y} = \mathcal{A}^{y}(X,Y), \nonumber\\
A^{z} &=& \pm \mathcal{A}^{t},
\end{eqnarray}
where the transformed coordinates, $X$ and $Y$, are given by
\begin{eqnarray} \label{XandY}
X = x - \psi(t \pm z), \ \ \ Y = y - \chi(t \pm z).
\end{eqnarray}
The functions $\psi$ and $\chi$ correspond to traveling waves and the choice of sign for $A^{z}$ must match the choice of sign in (\ref{XandY}). 
\footnote{The traveling wave functions, $\psi$ and $\chi$, used here are not to be confused with the angular coordinates in the compact space of the KS geometry, discussed in Sect. \ref{Wound-strings in the KS geometry}, which are denoted by the same Greek letters. In this appendix we follow the notation of Vachaspati and T. Vachaspati \cite{Vachaspati&Vachaspati} for the sake of clarity when consulting their original paper.}
It is straightforward to check that, if $\Phi(x,y)$ and $\mathcal{A}^{\mu}(x,y)$ satisfy the abelian-Higgs EOM (\ref{cov_EOMs})-(\ref{eq:EOM_vector_Abelian-Higgs_L}), the ansatz (\ref{Vachaspati_ansatz}) is also a valid solution. However, such a solution represents a string (of finite width), whose central-axis is located at the ``origin" with respect to the transformed coordinates
\begin{eqnarray} \label{axis_embedding}
X = Y = 0,
\end{eqnarray}
rather than at the origin of the physical coordinates of the background space, $x=y=0$. In the zero-width approximation (and switching from embedding to WS coordinates), these results reproduce those for the Nambu-Goto string outlined above. The Hamiltonian for the traveling wave solution is
\begin{eqnarray} \label{TW_Hamiltonian}
E = E_{LS} + \int d^3x \left\{|\dot{\psi}D_{X}\Phi + \dot{\chi}D_{Y}\Phi|^2 + \frac{B^2}{4\pi}(\dot{\psi}^2+\dot{\chi}^2)\right\},
\end{eqnarray}
where  $B \equiv \partial^{X}\mathcal{A}^{Y} - \partial^{Y}\mathcal{A}^{X}$, $D_{X} \equiv \partial_{X} - ie \mathcal{A}_{X}$, $D_{Y} \equiv \partial_{Y} - ie \mathcal{A}_{Y}$ and $E_{LS}$ is the energy of a long string in the absence of additional waves \cite{Vachaspati&Vachaspati}. Interestingly, using $E_{LS} \approx 2\pi \eta^2 |n| \Delta$ as in Sect. \ref{String energy and momentum}, and considering Nielsen-Olesen strings with a helical embedding in the background space, defined via
\begin{eqnarray} \label{Helical_Embedding}
X = x - r_0 \cos(\kappa(t-z)) = 0, \ \ \ Y = y - r_0 \sin(\kappa(t-z)) = 0, \ \ \ Z=z,
\end{eqnarray}
where $r_0 > r_{v|n|} \geq r_{s|n|}$ is the helix radius, the expression for the total energy takes a similar form to that given in (\ref{E_n}), i.e.
\begin{eqnarray} 
E \approx 2\pi \eta^2 |n|\Delta \left(1+\kappa^2 r_0^2\right).
\end{eqnarray}
However, here $r_0$ may take any value in the range $r_{v|n|} < r_0 < ct$, since it is not bounded by any physical consideration, besides causality, in ordinary Minkowski space. Furthermore, a helical configuration is not a very ``natural" configuration for a string in a non-compact space since, in order to form, it requires the application of constant centripetal force. By contrast, as discussed in Sect. \ref{Specific, but natural, models for winding formation}, helical configurations for the lines of constant phase and magnetic flux lines {\it within} the string may form if the scalar field takes a circuitous route from the top of the potential hill to the circle of degenerate minima. In this case, all that is required is an initial velocity perturbation at the phase-transition epoch. Likewise, in the case of wound $F$-strings, the periodic boundary conditions of the compact space ensure that a simple initial velocity perturbation at the ``end point" of the string (i.e. the point momentarily at the horizon), is sufficient to ensure dynamical winding formation.
\\ \indent
As mentioned in the introduction, in principle, a twisted flux-line string could, in addition, carry traveling waves described by a modified version of the usual formalism. For example, the helical embedding described by (\ref{Helical_Embedding}) could be applied also to the twisted flux-line solution, instead of to the Nielsen-Olesen string, as in \cite{Vachaspati&Vachaspati}. The string theory analogue of such a configuration would therefore be an $F$-string with a helical embedding in {\it both} four-dimensional Minkowski space and in the internal manifold.
\\ \indent
The traveling wave formalism was extended to strongly gravitating strings in \cite{Garfinkle&Vachaspati} by solving the full Einstein-abelian-Higgs EOM, and an equivalent extension for the twisted flux-line solution is clearly also worthwhile. However, perhaps an even more interesting approach would be to try first to develop a general ``traveling wave" type formalism for waves {\it within} strings (either in the EM field, or in the the phase factor and/or VEV of the scalar field). This is especially pertinent with regard to the analogy with wound $F$-strings. Since the embedding of left and right movers for the $F$-string in the compact space is arbitrary, it is crucial to determine if the embedding of magnetic flux lines and lines of constant phase within the vortex core of local strings, allowed by the abelian-Higgs equations,  is likewise. Such work would determine if the analogy between $F$-strings in extra-dimensions and gauge strings in four-dimensions holds for arbitrary motion of the former in the internal space.
\begin{center}
{\bf Acknowledgments}
\end{center}
ML was supported by JSPS Research Fellowship and Grant-in-Aid No.~P10327. JY was supported by the Grant-in-Aid for Scientific Research No.~23340058 and the Grant-in-Aid for Scientific Research on Innovative Areas No.~21111006. We would like to thank Jaume Garriga for his helpful comments.
\end{document}